\documentclass{aastex631}

\usepackage{siunitx}
\usepackage{physics}
\usepackage{graphicx}
\usepackage{makecell}
\usepackage{multirow}
\usepackage{CJK}
\usepackage[caption=false]{subfig}
\DeclareSIUnit\jansky{Jy}
\DeclareSIUnit\erg{erg}
\DeclareSIUnit\photon{photon}
\DeclareSIUnit\parsec{pc}
\DeclareSIUnit\year{yr}
\DeclareSIUnit\Msun{M_{\odot}}
\DeclareSIUnit\magnitude{mag}
\newcolumntype{I}{>{\hangindent=1em\hangafter=1}p{8cm}}

\shorttitle{GRB supervised ML}
\shortauthors{J-W Luo et al.}

\begin{document}
\title{Identifying the physical origin of gamma-ray bursts with supervised machine learning}
\author[0000-0002-9642-9682]{Jia-Wei Luo}
\affiliation{College of Physics and Hebei Key Laboratory of Photophysics Research and Application, Hebei Normal University, Shijiazhuang, Hebei 050024, China}
\affiliation{Nevada Center for Astrophysics, University of Nevada, Las Vegas, NV 89154, USA}
\affiliation{Department of Physics and Astronomy, University of Nevada Las Vegas, NV 89154, USA}

\author{Fei-Fei Wang}
\affiliation{School of Mathematics and Physics, Qingdao University of Science and Technology, Qingdao 266061, China}

\author{Jia-Ming Zhu-Ge}
\affiliation{Nevada Center for Astrophysics, University of Nevada, Las Vegas, NV 89154, USA}
\affiliation{Department of Physics and Astronomy, University of Nevada Las Vegas, NV 89154, USA}

\author{Ye Li}
\affiliation{Purple Mountain Observatory, Chinese Academy of Sciences, Nanjing 100012, China}

\author[0000-0002-5400-3261]{Yuan-Chuan Zou}
\affiliation{Department of Astronomy, School of Physics, Huazhong University of Science and Technology, Wuhan, 430074, People's Republic of China}

\author[0000-0002-9725-2524]{Bing Zhang}
\affiliation{Nevada Center for Astrophysics, University of Nevada, Las Vegas, NV 89154, USA}
\affiliation{Department of Physics and Astronomy, University of Nevada Las Vegas, NV 89154, USA}

\correspondingauthor{Jia-Wei Luo}
\email{ljw@hebtu.edu.cn}
\begin{abstract}
The empirical classification of gamma-ray bursts (GRBs) into long and short GRBs based on their durations is already firmly established. This empirical classification is generally linked to the physical classification of GRBs originating from compact binary mergers and GRBs originating from massive star collapses, or Type I and II GRBs, with the majority of short GRBs belonging to Type I and the majority of long GRBs belonging to Type II. However, there is a significant overlap in the duration distributions of long and short GRBs. Furthermore, some intermingled GRBs, i.e., short-duration Type II and long-duration Type I GRBs, have been reported. A multi-parameter classification scheme of GRBs is evidently needed. In this paper, we seek to build such a classification scheme with supervised machine learning methods, chiefly \textsc{XGBoost}. We utilize the GRB Big Table and Greiner's GRB catalog and divide the input features into three subgroups: prompt emission, afterglow, and host galaxy. We find that the prompt emission subgroup performs the best in distinguishing between Type I and II GRBs. We also find the most important distinguishing feature in prompt emission to be $T_{90}$, hardness ratio, and fluence. After building the machine learning model, we apply it to the currently unclassified GRBs to predict their probabilities of being either GRB class, and we assign the most probable class of each GRB to be its possible physical class.
\end{abstract}


\section{Introduction}
\label{sec:introduction}
Dating from the early days of gamma-ray burst (GRB) study, a clear bimodal distribution had been identified in their durations \citep{kouveliotou1993IdentificationTwoClasses}. Two classes of GRBs are then proposed based on their durations, namely long GRBs (LGRBs) and short GRBs (SGRBs). The commonly used criterion is based on $T_{90}$, the time within which 90\% of the fluence of the GRB is observed, with the dividing point set to be $T_{90}=\SI{2}{\s}$. 

LGRBs are thought to be produced by the core-collapse of massive stars \citep{woosley1993GammarayBurstsStellar}, and this theory is subsequently supported by direct observational evidence of the association of some LGRBs with Type Ic supernovae \citep{galama1998UnusualSupernovaError, woosley2006SupernovaGammarayBurst}. SGRBs are thought to be originated from compact star mergers \citep{eichler1989NucleosynthesisNeutrinoBursts}, and this theory is supported by the multi-messenger observations of the binary neutron star merger event GW170817/GRB 170817A \citep{abbott2017GW170817ObservationGravitational,abbott2017MultimessengerObservationsBinary,abbott2017GravitationalWavesGammaRays,goldstein2017OrdinaryShortGammaRay,zhang2018PeculiarLowluminosityShort}.

However, this dichotomy is far from perfect. Significant overlap presents in the duration distributions of long and short GRBs, and the duration itself is dependent on the energy band in which it is measured \citep{mukherjee1998ThreeTypesGamma,hakkila2003HowSampleCompleteness,horvath2006NewDefinitionIntermediate,zhang2008AnalysisDurationsSwift,veres2010DISTINCTPEAKFLUXDISTRIBUTION,qin2012COMPREHENSIVEANALYSISFERMI,bromberg2013SHORTLONGCOLLAPSARS,zhang2016GRBObservationalProperties}. Moreover, there are some short-duration GRBs thought to be possibly produced by core-collapse massive stars \citep{greiner2009GRB080913REDSHIFT,tanvir2009GrayBurstRedshift,salvaterra2009GRB090423Redshift,antonelli2009GRB090426Farthest,zhang2009DISCERNINGPHYSICALORIGINS,guelbenzu2011GRB090426Discovery,zhang2021PeculiarlyShortdurationGammaray,ahumada2021DiscoveryConfirmationShortest,rossi2022PeculiarShortdurationGRB}, as well as some long-duration GRBs thought to be possibly originated from compact star mergers \citep{gal-yam2006NovelExplosiveProcess,gehrels2006NewGrayBurst,fynbo2006NoSupernovaeAssociated,dellavalle2006EnigmaticLonglastingGray,zhang2007MakingShortGammaRay,troja2022NearbyLongGammaray,yang2022LongdurationGammarayBurst,rastinejad2022KilonovaFollowingLongduration,sun2023MagnetarEmergencePeculiar}.

The existence of these ``intermingled'' GRBs challenges the practice of classifying GRBs solely based on duration, as well as the names of ``long" and ``short" GRBs. It is then apparent that more sophisticated classification criteria involving multiple observational parameters are needed. Throughout this study, we refer to the GRB classes based on their physical origins, namely Type I for compact merger GRBs, and Type II for collapsar GRBs, following the classification scheme of \citet{zhang2006BurstNewIdeas,zhang2009DISCERNINGPHYSICALORIGINS}. Many other schemes for GRB classification have also been put forward \citep[e.g.][]{zhang2009DISCERNINGPHYSICALORIGINS,lu2010NEWCLASSIFICATIONMETHOD,zhang2012RevisitingLongSoftShort,bromberg2013SHORTLONGCOLLAPSARS,lu2014AmplitudeParameterGammaray,yang2016TwoDimensionalClassification,li2016ComparativeStudyLong,kulkarni2017ClassificationGammarayBurst,li2020ComparativeStudyLong,minaev2020EpEisoCorrelation}, yet the classification of long and short GRBs is still largely based on community consensus, and there is a lack of objective classification models with minimal human interference.

In this case, machine learning comes in handy. Capable of automatically generating results without human input after training, machine learning can help us to fathom the differences between Type I and II GRBs, as well as aid us in the classification of newly discovered GRBs. Machine learning have already been widely adopted in the study of GRBs \citep[e.g.][]{horvath2006NewDefinitionIntermediate,ripa2012SPECTRALLAGSPEAK,huertas-company2015CATALOGVISUALLIKEMORPHOLOGIES,tarnopolski2015DistinguishingShortLong,modak2018ClusteringGammarayBursts,horvath2019MultidimensionalAnalysisFermi,jespersen2020UnambiguousSeparationGammaRay,salmon2022TwoDimensionalClustering,modak2021DistinctionGroupsGammaray,salmon2022TwoClassesGammaray,tarnopolski2022GraphbasedClusteringGammaray,bhave2022TwoDimensionalClustering,steinhardt2023ClassificationBATSESwift}. However, the above-mentioned studies predominantly use machine learning methods of the unsupervised type, where only the observed features of the GRBs are inputted into the models, but not the labels (the GRBs' physical classes being Type I or II). On the other hand, the other type of machine learning methods, supervised methods, are also commonly employed by astronomy researchers in the classification of other astronomical objects \citep[e.g.][]{connor2018ApplyingDeepLearning,villa-ortega2022AstrophysicalSourceClassification,butter2022ClassificationFermiLATBlazars,beurs2022ComparativeStudyMachinelearning,yang2022ClassifyingUnidentifiedXRay,coronado-blazquez2022ClassificationFermiLATUnidentified,kaur2023UsingNeuralNetworks,fan2022ClassificationBlazarCandidates,luo2023MachineLearningClassification,zhu-ge2023MachineLearningClassification}, albiet study on the application of supervised methods on GRB is scarce. Since supervised methods take both features and labels as input, and can produce deterministic predictions of the class of new GRBs, they can be helpful in identifying the true physical origin of intermingled GRBs.

In this study, we apply supervised machine learning methods to the classification of Type I and II GRBs. The machine learning model we use is the eXtreme Gradient Boosting (\textsc{XGBoost}) classifier \citep{chen2016XGBoostScalableTree}. We employ \textsc{XGBoost} as it is one of the most popular and successful machine learning frameworks to date, and can handle missing input values natively. The GRB catalog we use contains many missing values, so the ability to handle them is vital. In Section \ref{sec:data}, we introduce the GRB catalogs we utilize and the machine learning methods we use. In Section \ref{sec:results}, we present the classification results and feature importance from the machine learning models. In Section \ref{sec:prediction}, we attempt to predict the classes of the unclassified GRBs. Finally, in Section \ref{sec:conclusions}, we put forward our conclusions and discuss on the classifications of some recently discovered possible intermingled GRBs.

\section{Data and methods}
\label{sec:data}

We use an updated version of the GRB Big Table (\citet{wang2020ComprehensiveStatisticalStudy}, Wang et al. (in prep)), which contains 7179 GRBs ranging from 1991 April 21 -- 2021 July 08. Greiner's GRB catalog (\url{https://www.mpe.mpg.de/~jcg/grbgen.html}), on the other hand, has 2261 GRBs in the same time range. We match the two catalogs, requiring $T_{90}$ of the selected GRBs in the Big Table to be known, and we label the GRBs based on their labels in Greiner's catalog. GRBs with `S' at the end of their names are marked as Type I GRBs, while the others are marked as Type II GRBs. We also adopt the consensus classification of some intermingled GRBs: Type II GRB 090426 \citep{antonelli2009GRB090426Farthest,guelbenzu2011GRB090426Discovery}, Type I GRB 060505 \citep{fynbo2006NoSupernovaeAssociated} and Type I GRB 060614 \citep{fynbo2006NoSupernovaeAssociated,gal-yam2006NovelExplosiveProcess,gehrels2006NewGrayBurst,zhang2007MakingShortGammaRay}. This leaves us with 144 Type I and 1761 Type II GRBs. We acknowledge that this matching method substantially reduces the size of our sample, but the unmatched GRBs do not have many known features, to begin with. Therefore, we did not discard too much information.

The classification input of our model is based on the Greiner's catalog, which collected the community consensus based on both $T_{90}$ and afterglow/host galaxy information as presented in the literature. It is possible that a small fraction of bursts is mis-classified, but the very strength of our machine learning model is that it considers all the classifications of the input training sample. If there are a few GRBs that are wrongly classified, they would not have a significant impact on the overall accuracy.

In this study, we pay special interest to the intermingled GRBs. We define intermingled GRBs as GRBs classified as Type I in Greiner's catalog, but have $T_{90}$ values $>\SI{2}{\s}$ in the Big Table, or GRBs classified as Type II in Greiner's catalog, but have $T_{90}<\SI{2}{\s}$. There are 21 intermingled Type I GRBs and 59 intermingled Type II GRBs in our sample.

We also scrutinize the possible third intermediate GRB type proposed by some studies \citep[e.g.][]{horvath1998ThirdClassGamma,mukherjee1998ThreeTypesGamma,hakkila2000GammaRayBurstClass,balastegui2001ReclassificationGammarayBursts,hakkila2003HowSampleCompleteness,horvath2006NewDefinitionIntermediate,chattopadhyay2007StatisticalEvidenceThree,horvath2008ClassificationSwiftGammaray,huja2009ComparisonGammarayBursts,ripa2009SearchGammarayBurst,veres2010DISTINCTPEAKFLUXDISTRIBUTION,horvath2010DETAILEDCLASSIFICATIONLess,ripa2012SPECTRALLAGSPEAK,koen2012MultipleClassesGammaray,zitouni2015StatisticalStudyObserved,kulkarni2017ClassificationGammarayBurst,horvath2018ClassifyingGRB170817A} based on the $T_{90}$ distributions of GRBs by creating an intermediate sample consisting of GRBs with $T_{90}$ within $\SIrange{1}{4}{\s}$. There are 31 intermediate Type I and 100 intermediate Type II GRBs in our sample.

We then divide the features in the Big Tables into three subgroups: prompt emission, afterglow and host galaxy. Three subsamples are subsequently created by requiring each GRB in the subsamples to have at least one feature other than $T_{90}$ in the corresponding feature group to be known. We also divide each subsample into training sets and test sets with a 7:3 ratio, while keeping the ratio of Type I to Type II GRBs the same in the training sets and test sets. The training sets are used to train the machine learning model, while the test sets are used to test the performance of the model after it is trained. 

\begin{table*}
    \centering
    \begin{tabular}{llll}
    \hline
    Feature name & Unit & Description & Log \\
	\hline
	\texttt{T90} & $\si{\s}$ & Time within which 90\% of the fluence of the GRB is observed & Y\\
    \texttt{F\_g} & $10^{-6}\;\si{\erg\per\centi\m\squared}$ & Fluence in the $\SIrange{20}{2000}{\kilo\eV}$ energy band & Y\\
    \texttt{HR} &  --- & Hardness ratio between $\SIrange{100}{2000}{\kilo\eV}$ and $\SIrange{20}{100}{\kilo\eV}$& Y\\
    \texttt{F\_pk1} & $10^{-6}\;\si{\erg\per\centi\m\squared\per\s}$ & Peak flux in the \SI{1}{\s} time bin in the rest-frame \SIrange{1}{10e4}{\kilo\eV} energy band & Y\\
    \texttt{P\_pk4} & \si{\photon\per\centi\m\squared\per\s} & Peak photon flux in the \SI{1}{\s} time bin of \SIrange{10}{1000}{\kilo\eV} & Y\\
    \texttt{alpha\_band} & --- & Low-energy spectrum index of the Band model & N\\
    \texttt{beta\_band} & --- & High-energy spectrum index of the Band model & N\\
    \texttt{E\_P\_band} & $\si{\kilo\eV}$ & Spectral peak energy of the Band model & Y\\
    \texttt{alpha\_cpl} & --- & Spectrum index of the cutoff power-law (CPL) model & N\\
    \texttt{E\_P\_cpl} & $\si{\kilo\eV}$ & Spectral peak energy of the cutoff power-law (CPL) model & Y\\
    \texttt{alpha\_spl} & --- & Spectrum index of the simple power-law (SPL) model & N\\
    \texttt{spectral\_lag} & \si{\milli\s\per\mega\eV} & Spectral time lag & N\\
    \hline
    \texttt{z} & --- & Redshift & N\\
    \texttt{D\_L} & $10^{28}\;\mathrm{cm}$ & Luminosity distance & Y\\
    \texttt{E\_iso} & $10^{52}\;\si{\erg}$ & Isotropic gamma-ray energy in the rest-frame \SIrange{1}{10e4}{\kilo\eV} energy band & Y\\
    \texttt{L\_pk} & $10^{52}\;\si{\erg\per\s}$ & Isotropic peak luminosity in the \SI{1}{\s} time bin in the rest-frame \SIrange{1}{10e4}{\kilo\eV} energy band & Y\\
	\hline
	\end{tabular}
	\caption{List of features used in the prompt emission subgroup. For features with multiple definitions (e.g., variability, F\_pk), we choose the one with the most known data. Directly measured features are listed above the horizontal line, while the derived features are listed below the line.}
	\label{table:list_features_prompt}
\end{table*}

\begin{table*}
    \centering
    \begin{tabular}{llll}
    \hline
    Feature name & Unit & Description & Log \\
	\hline
    \texttt{theta\_j} & \si{\radian} & Jet-opening angle & Y\\
    \texttt{Gamma0} & --- & Initial Lorentz factor & Y\\
    \texttt{log\_t\_burst} & \si{\s} & Duration of the GRB central engine & Y\\
    \texttt{t\_b} & \si{\day} & Jet break time & Y\\
    \texttt{F\_X11hr} & \si{\jansky} & Flux density in the X-ray band \SI{11}{\hour} after the trigger time of the burst & Y\\
    \texttt{beta\_X11hr} & --- & Index in X-ray band \SI{11}{\hour} after the trigger time of the burst & N\\
    \texttt{F\_Opt11hr} &  \si{\jansky} & Flux density in the optical band \SI{11}{\hour} after the trigger time of the burst & Y\\
    \texttt{T\_ai} & \si{\s} & Rest-frame time at the end of the plateau phase in log in X-ray & Y\\
    \texttt{L\_a} & \si{\erg\per\s} & Isotropic X-ray luminosity at the time Ta & Y\\
	\hline
	\end{tabular}
	\caption{List of features used in the afterglow subgroup.}
	\label{table:list_features_afterglow}
\end{table*}

\begin{table*}
    \centering
    \begin{tabular}{llll}
    \hline
    Feature name & Unit & Description & Log \\
	\hline
    \texttt{offset} & \si{\kilo\parsec} & Distance from the burst location to the center of the host galaxy & Y\\
    \texttt{metallicity} & --- & Metallicity of the host; the value is 12 + log[O/H] & N\\
    \texttt{A\_V} & --- & Dust extinction & N\\
    \texttt{SFR} & $\si{\Msun\per\year}$& Star formation rate & Y\\
    \texttt{SSFR} & \si{\per\giga\year} & Specific star formation rate & Y\\
    \texttt{Age} & \si{\mega\year} & The age of the GRB host galaxy & Y\\
    \texttt{Mass} & $\si{\Msun}$ & Stellar mass & Y\\
	\hline
	\end{tabular}
	\caption{List of features used in the host galaxy subgroup.}
	\label{table:list_features_host}
\end{table*}

While it is common practice to impute the missing values in the data with some type of algorithm, we find that imputation introduces false information in the feature importance we later calculate, which is also suggested by some other studies \citep[e.g.][]{seijo-pardo2019BiasesFeatureSelection,yu2022CausalFeatureSelection}. Since the \textsc{XGBoost} classifier \citep{chen2016XGBoostScalableTree} can automatically handle missing values, we simply input our data without imputation.

Then, we note that the Type I and Type II GRBs in our sample are significantly imbalanced by a ratio of $\sim 1:10$. Because this apparent ratio could be caused by selection effects, we should not introduce this ratio to our training data. However, the commonly used synthetic minority over-sampling technique (SMOTE) \citep{chawla2002SMOTESyntheticMinority} cannot be applied to data with missing values. Instead, we assign different sample weights for the two classes calculated with a balanced sample weight implemented in \textsc{scikit-learn} \citep{pedregosa2011ScikitlearnMachineLearning}:
\begin{equation}
    w_i = \frac{N}{kn_i},
\end{equation}
where $w_i$ is the sample weight of the $i$th class, $N$ is the total number of data points, $k$ is the number of classes (in this study 2), and $n_i$ is the number of data points in the $i$th class.

Finally, we input the training sets into the \textsc{XGBoost} classifier to train the machine learning model. After training, we use the test set and the commonly used $F_1$ score \citep{vanrijsbergen1979InformationRetrieval,sasaki2007TruthFmeasure} to assess the performance of our models. A more intuitive metric, accuracy, is disfavored here because our data is imbalanced. A model simply predicts all GRBs as Type II can still score 92\% accuracy.

To calculate the $F_1$ score, we first consider two commonly used metrics in evaluating the performance of machine learning models, precision and recall:
\begin{itemize}
    \item Precision
    \begin{equation}
        \mathrm{Precision} = \frac{\mathrm{True\;positives}}{\mathrm{True\;positives} + \mathrm{False\;positives}}.
    \end{equation}
    Precision measures how many of the items predicted by the model as positive (in this study Type I GRBs) are true positives.
    
    \item Recall
    \begin{equation}
        \mathrm{Recall} = \frac{\mathrm{True\;positives}}{\mathrm{True\;positives} + \mathrm{False\;negatives}}.
    \end{equation}
    Recall measures how many of the originally positive items are correctly predicted as positive by the model.
\end{itemize}

$F_1$ score is then calculated as the harmonic mean of the two metrics:
\begin{equation}
    F_1 = 2\frac{\mathrm{Precision}\cdot\mathrm{Recall}}{\mathrm{Precision}+\mathrm{Recall}}.
\end{equation}

The resulting $F_1$ score is a value between $\SIrange{0}{1}{}$, with $0$ meaning total failure in predicting the correct labels for the test set, and $1$ meaning 100\% accuracy in predicting the labels. Since $F_1$ score is a stricter metric than accuracy, $F_1$ scores are usually significantly smaller than accuracy scores calculated on the same classification results.

To test which input feature has the best capability in distinguishing between Type I and II GRBs, We use SHapley Additive exPlanations (SHAP, \citet{lundberg2017UnifiedApproachInterpreting,lundberg2020LocalExplanationsGlobal}) to calculate the feature importance of the input features. For each data point, SHAP estimates the contribution towards the output result from each input feature in the form of SHAP values. Readers can refer to the above-mentioned references for more detailed and mathematical description of SHAP. In contrast to the also commonly used permutation feature importance \citep{breiman2001RandomForests,altmann2010PermutationImportanceCorrected,fisher2019AllModelsAre}, which generates a single feature importance value for a feature across all data points, SHAP can analyze the prediction contribution of features on individual data points. When the SHAP values from all the data points are combined, SHAP can show not only the importance of the input features, but also in which direction the feature values of each input data point draw the final output.

Figure \ref{fig:prompt_fi} and \ref{fig:prompt_fi2} show an example of the results from \textsc{SHAP}. In Figure \ref{fig:prompt_fi}, the SHAP values from each individual data points are taken absolute value and averaged across different features. The length of each bar in the figure shows how important is each feature to the prediction result in general.  Figure \ref{fig:prompt_fi2}, on the other hand, shows the individual SHAP values of each feature in each data points. In this beeswarm plot, the $X$-axis shows the SHAP values, with higher SHAP values leaning toward Type I, and lower SHAP values leaning toward Type II. The feature values of each data point are also shown with the color of the points, so that readers can know in which direction a higher or lower value in one feature draws the prediction to.

\section{Results}
\label{sec:results}
\subsection{Prompt emission}
\label{subsec:prompt_emission}

Many studies suggested adding hardness ratio (HR) to the $T_{90}$ classification criterion to form a two-dimensional criteria will yield better results \citep[e.g.][]{horvath2006NewDefinitionIntermediate,horvath2010DETAILEDCLASSIFICATIONLess,ripa2012SPECTRALLAGSPEAK,zhang2012RevisitingLongSoftShort,bhat2016THIRDFERMIGBM,yang2016TwoDimensionalClassification,horvath2018ClassifyingGRB170817A,tarnopolski2019AnalysisDurationHardness,zhang2022DistributionGammarayBursts}. Similarly, the power-law index or peak energy $E_p$ of the spectrum of prompt emission can also take the place of hardness ratio \citep{zhang2012RevisitingLongSoftShort,goldstein2010NEWDISCRIMINATORGAMMARAY,nava2011SpectralProperties438}. In general, Type I GRBs have harder spectra compared with Type II GRBs. \citet{goldstein2010NEWDISCRIMINATORGAMMARAY} further proposes classification on the $E_p$ -- fluence plane. Since fluence is highly related to duration, this scheme also follows the HR - $T_{90}$ scheme.

Some other studies \citep[e.g.][]{zhang2009DISCERNINGPHYSICALORIGINS,zhang2012RevisitingLongSoftShort,qin2013StatisticalClassificationGammaray,tsutsui2013PossibleExistenceEpLp,minaev2020EpEisoCorrelation} suggest that the famous Amati relation \citep{amati2002IntrinsicSpectraEnergetics,amati2009ExtremelyEnergeticFermi,kumar2015PhysicsGammarayBursts} of the peak energy $E_p$ and the isotropic energy $E_{iso}$ of GRB prompt emission are different for Type I and II GRBs, and thus the $E_p$ -- $E_{iso}$ plane can be used to distinguish between Type I and II GRBs.

In addition, \citet{norris2006ShortGammaRay,yi2006SpectralLagsShort,gehrels2006NewGrayBurst,zhang2006BurstNewIdeas,ukwatta2010SPECTRALLAGSLAGLUMINOSITY,minaev2014CatalogShortGammaray,bernardini2015ComparingSpectralLag,shao2017NewMeasurementSpectral} propose to classify Type I and II GRBs based on spectral lag $\tau$ and the $\tau$ -- peak luminosity $L_p$ plane, where Type I GRBs have smaller spectral lags and peak luminosities.

Association with supernovae (SN) is also a very important distinguishing factor between Type I and II GRBs, as SN associations provide smoking-gun evidence of the GRB progenitor. However, the Big Table only contains SN association information for 22 GRBs. For those GRBs without SN association information, it is unknown whether there truly was no SN associated with the GRB, or there simply was no observation, or most likely, there was an optical observation, but the SN was outshone by the bright optical afterglow. {We find that including SN association in our model results in significantly lower $F_1$ scores. While the model correctly classifies GRBs with SN detection as Type II GRBs, including SN in our model also makes it more likely to classify GRBs without SN detection as Type I. Since most GRBs do not have SN detection in our data because they are too far away for SN detection, and because there are more Type II GRBs than Type I, including SN will yield worse results. Furthermore, the model without SN can correctly classify almost all SN associated GRBs as Type II. Therefore, we do not include the SN information in our model.

\begin{figure*}
\centering
\subfloat[Confusion matrix on all GRBs in the test set]{
    \includegraphics[width=0.48\textwidth]{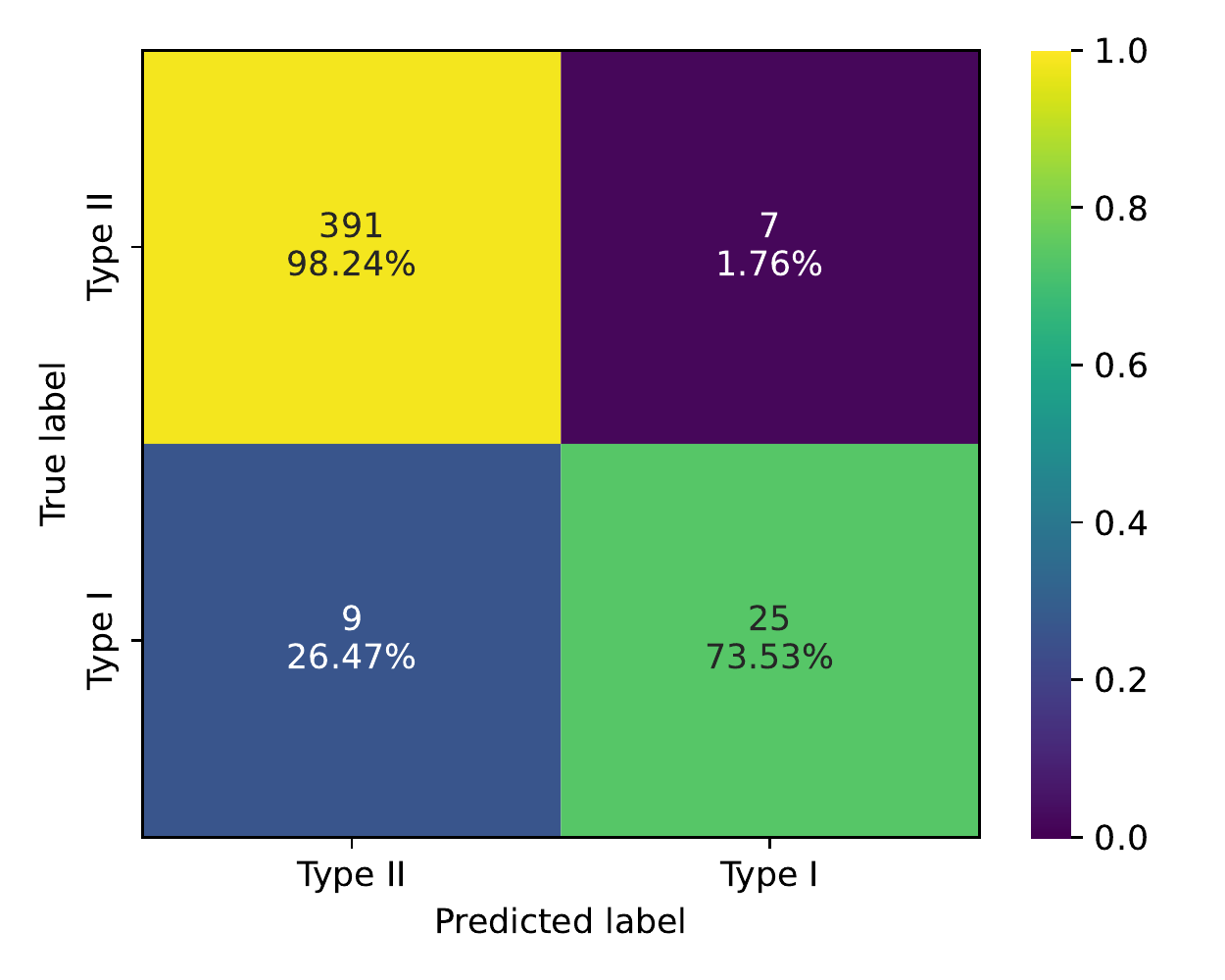}
    \label{fig:prompt_cm}}
\subfloat[Confusion matrix on intermingled GRBs in the test set]{
    \includegraphics[width=0.48\textwidth]{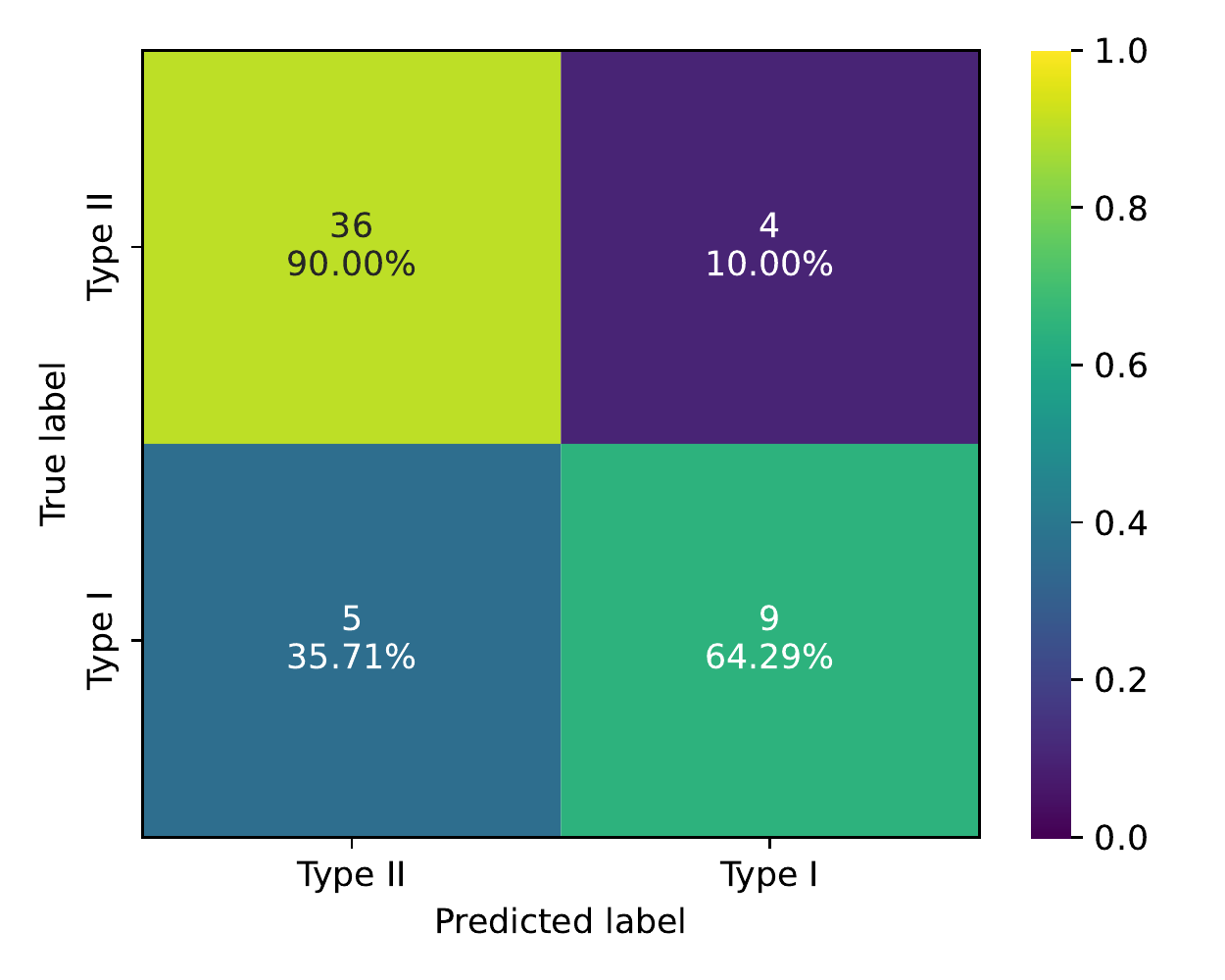}
    \label{fig:prompt_inter_cm}}\\
\subfloat[Confusion matrix on intermediate GRBs in the test set]{
    \includegraphics[width=0.48\textwidth]{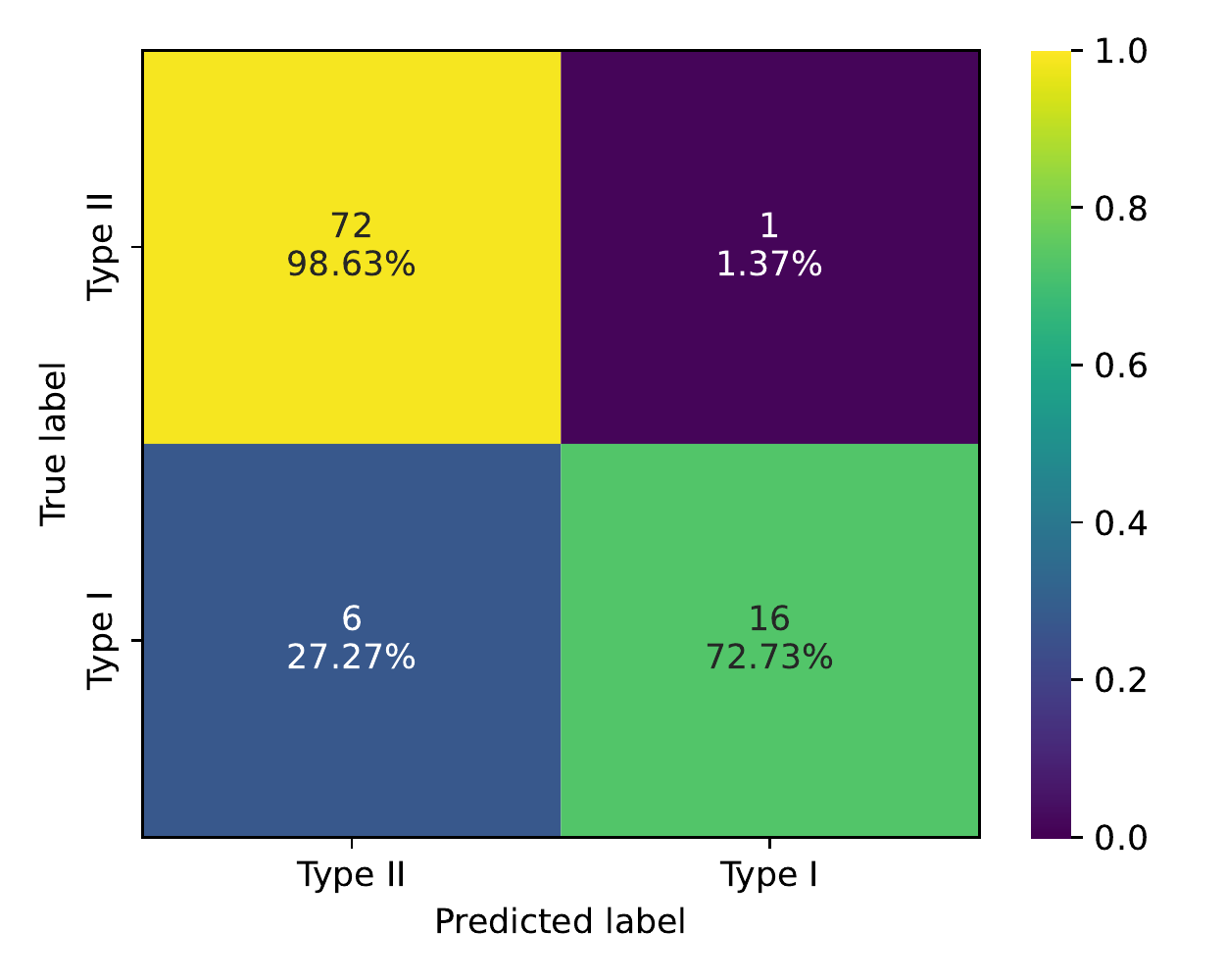}
    \label{fig:prompt_mediate_cm}}
\subfloat[Average SHAP values of each feature on the training set]{
    \includegraphics[width=0.48\textwidth]{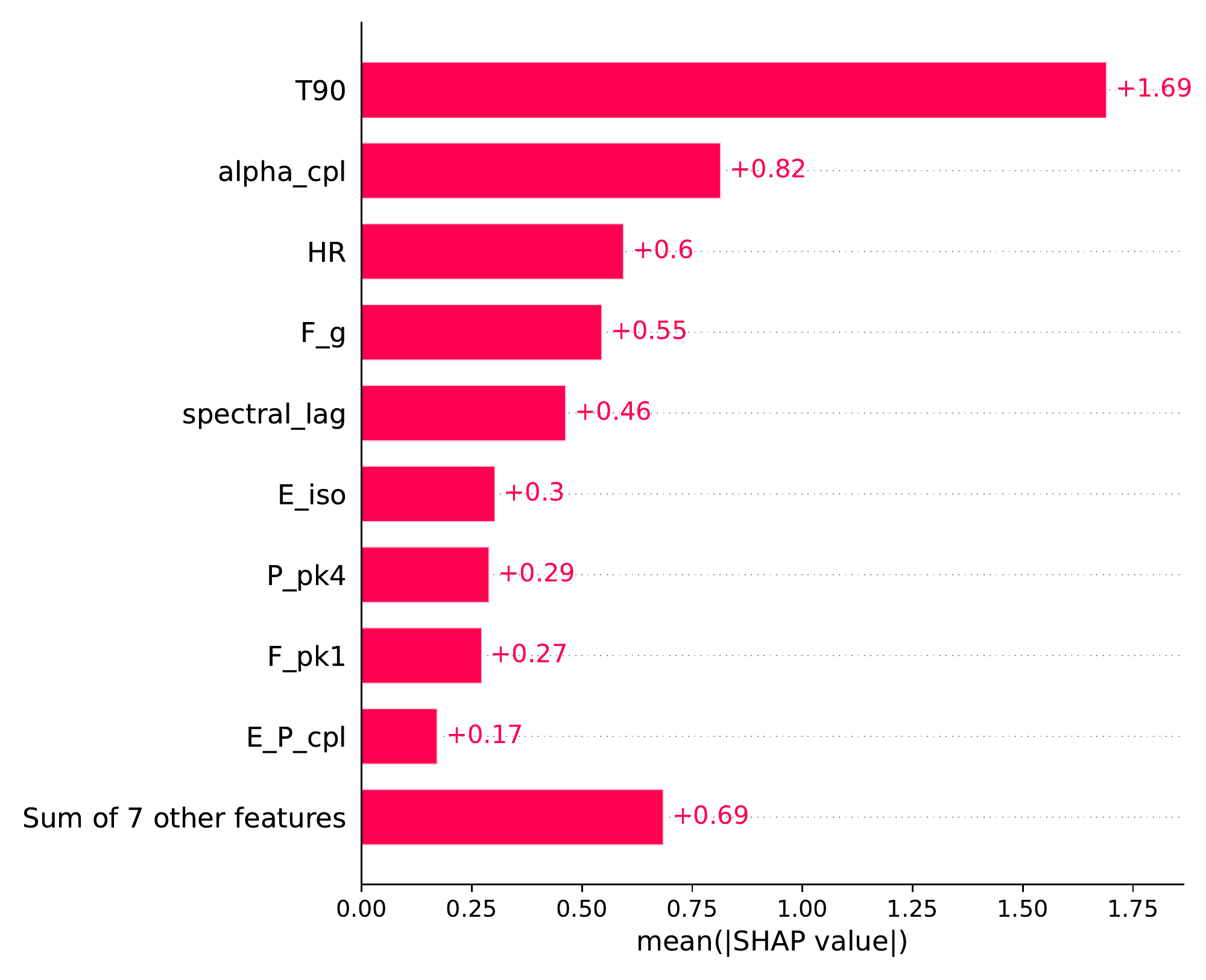}
    \label{fig:prompt_fi}}\\
\subfloat[SHAP value beeswarm plot on the training set]{
    \includegraphics[width=0.48\textwidth]{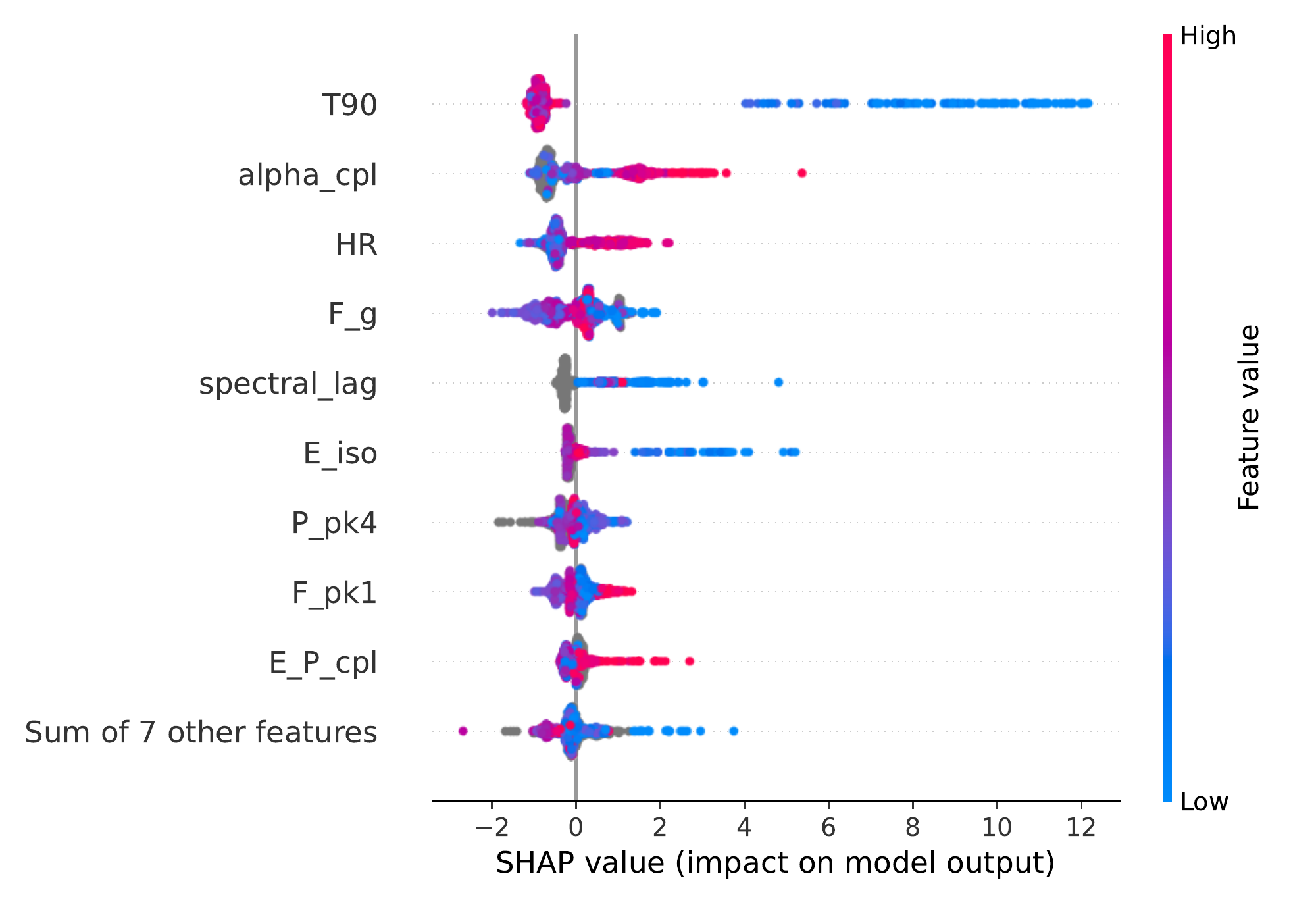}
    \label{fig:prompt_fi2}}
\caption{Examples of confusion matrices and SHAP feature importance values of the prompt emission subgroup.}
\label{fig:prompt}
\end{figure*}

With the prompt emission subgroup, we are able to obtain a $F_1$ score of 0.758 on the test set, 0.667 on the intermingled GRBs, and 0.821 on the intermediate GRBs. The corresponding confusion matrices and feature importance are shown in Figure \ref{fig:prompt}. Our model can predict most GRBs correctly based on prompt emission data, and $T_{90}$ is the most prominent feature, with feature importance much higher than other features, and shorter $T_{90}$ pull the predictions toward Type I. Since the intermingled GRBs are the ones that defy the classification based on $T_{90}$, the major features that cause their classifications to be different will be the features that have high feature importance other than T90. The same stands true for intermediate GRBs, if $T_{90}$ cannot classify them clearly, then they will be classified based on other important features.

\begin{figure*}
\centering
\subfloat[Confusion matrix on all GRBs in the test set]{
    \includegraphics[width=0.48\textwidth]{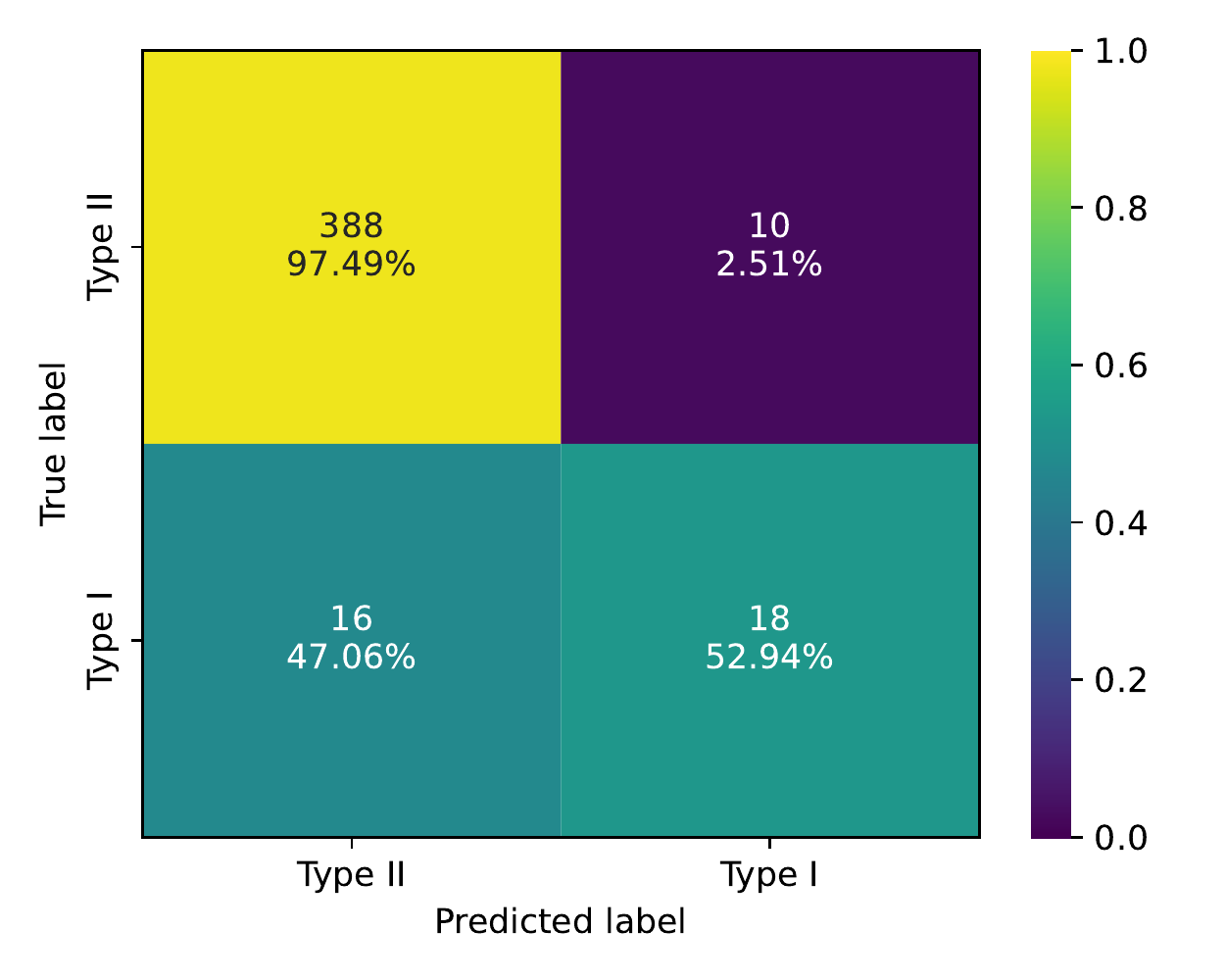}
    \label{fig:prompt_no_t90_cm}}
\subfloat[Confusion matrix on intermingled GRBs in the test set]{
    \includegraphics[width=0.48\textwidth]{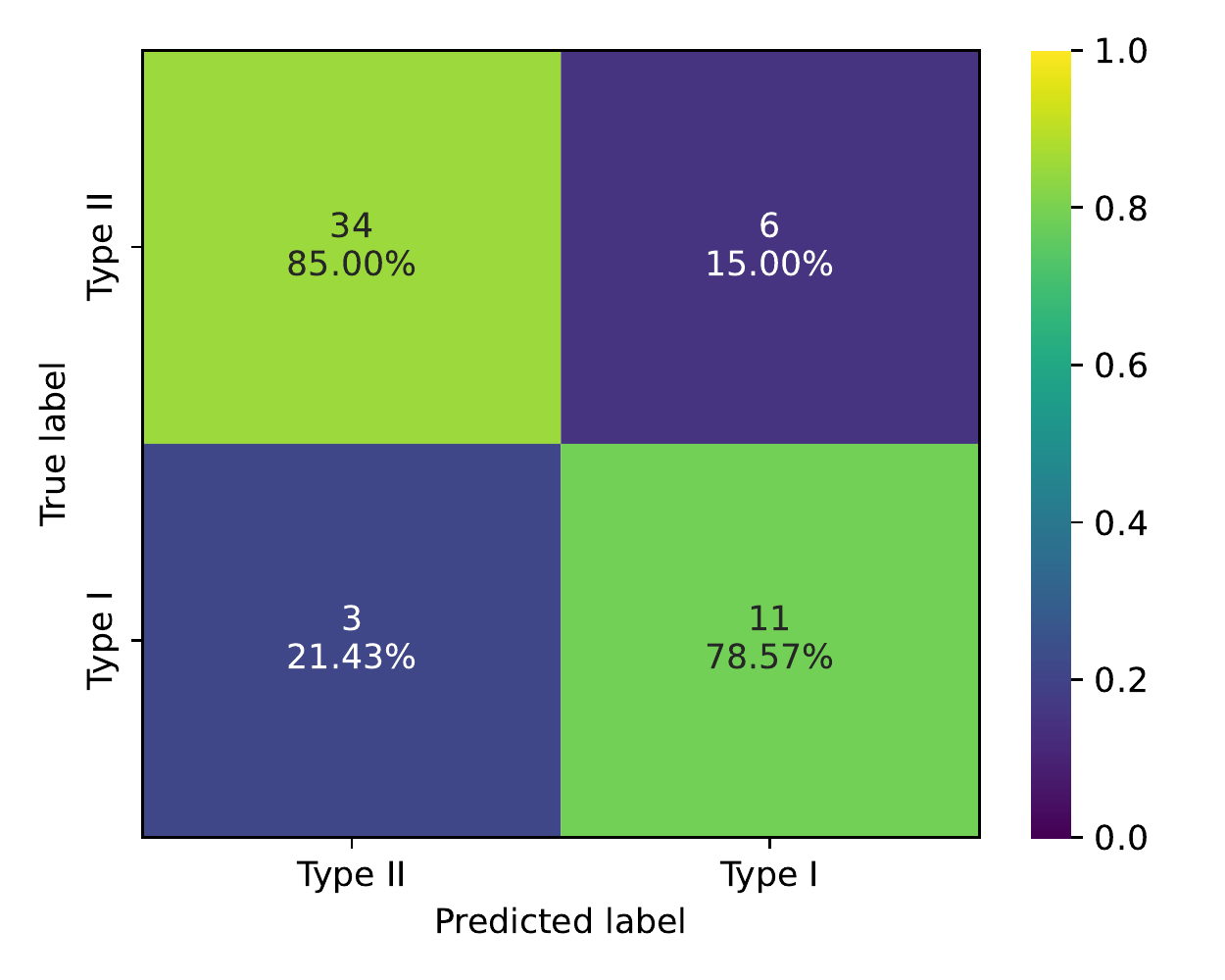}
    \label{fig:prompt_no_t90_inter_cm}}\\
\subfloat[Confusion matrix on intermediate GRBs in the test set]{
    \includegraphics[width=0.48\textwidth]{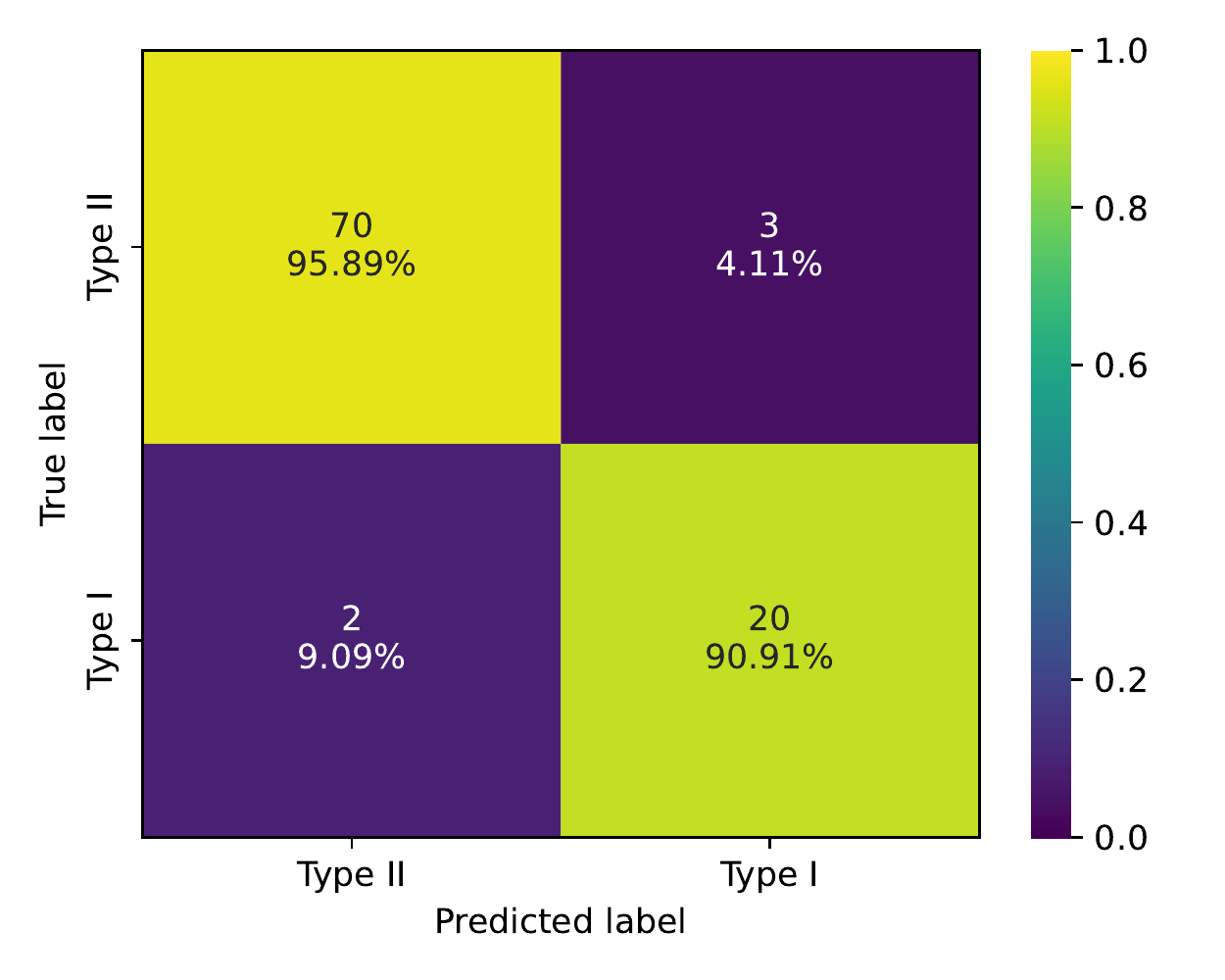}
    \label{fig:prompt_no_t90_mediate_cm}}
\subfloat[Average SHAP values of each feature on the training set]{
    \includegraphics[width=0.48\textwidth]{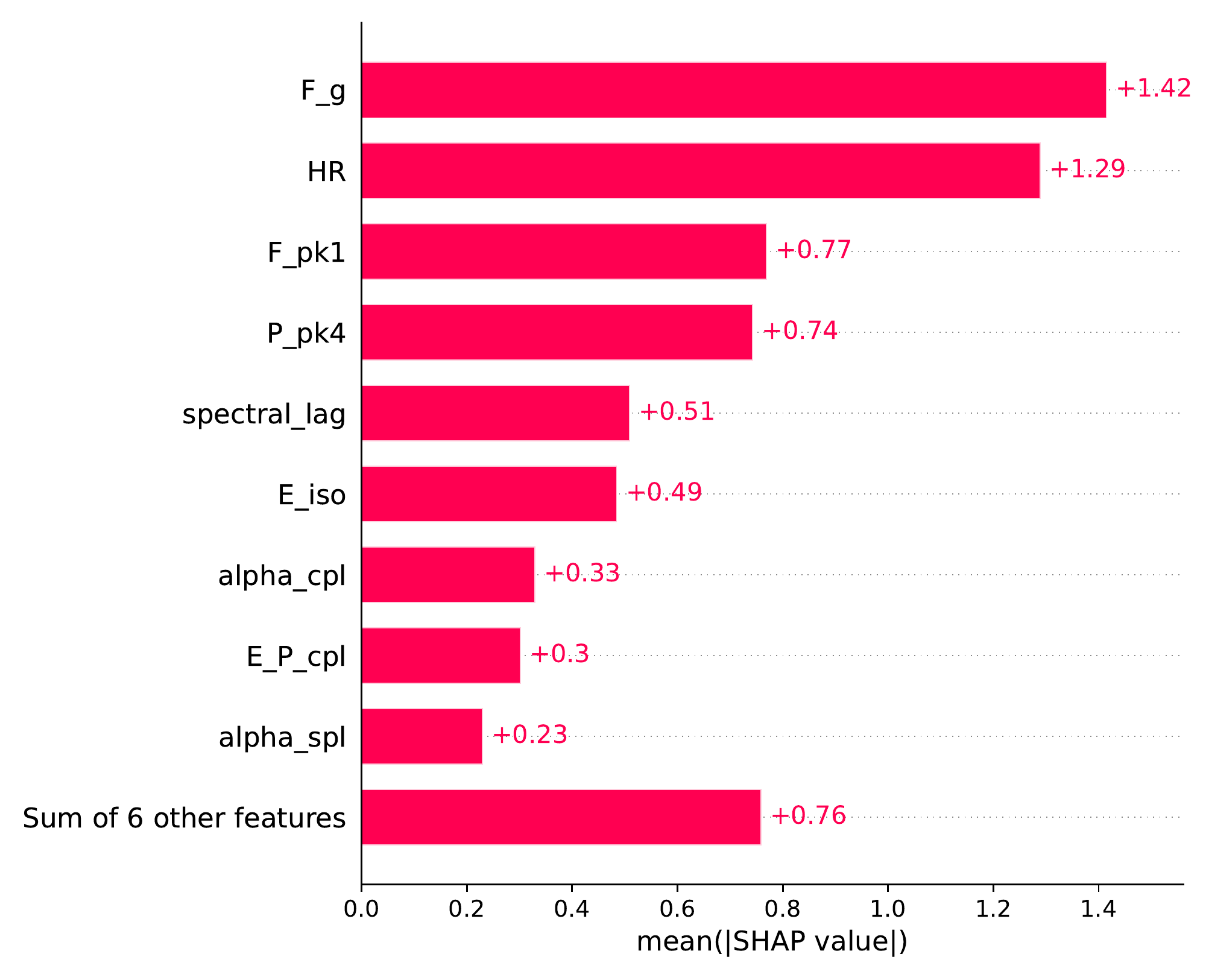}
    \label{fig:prompt_no_t90_fi}}\\
\subfloat[SHAP value beeswarm plot on the training set]{
    \includegraphics[width=0.48\textwidth]{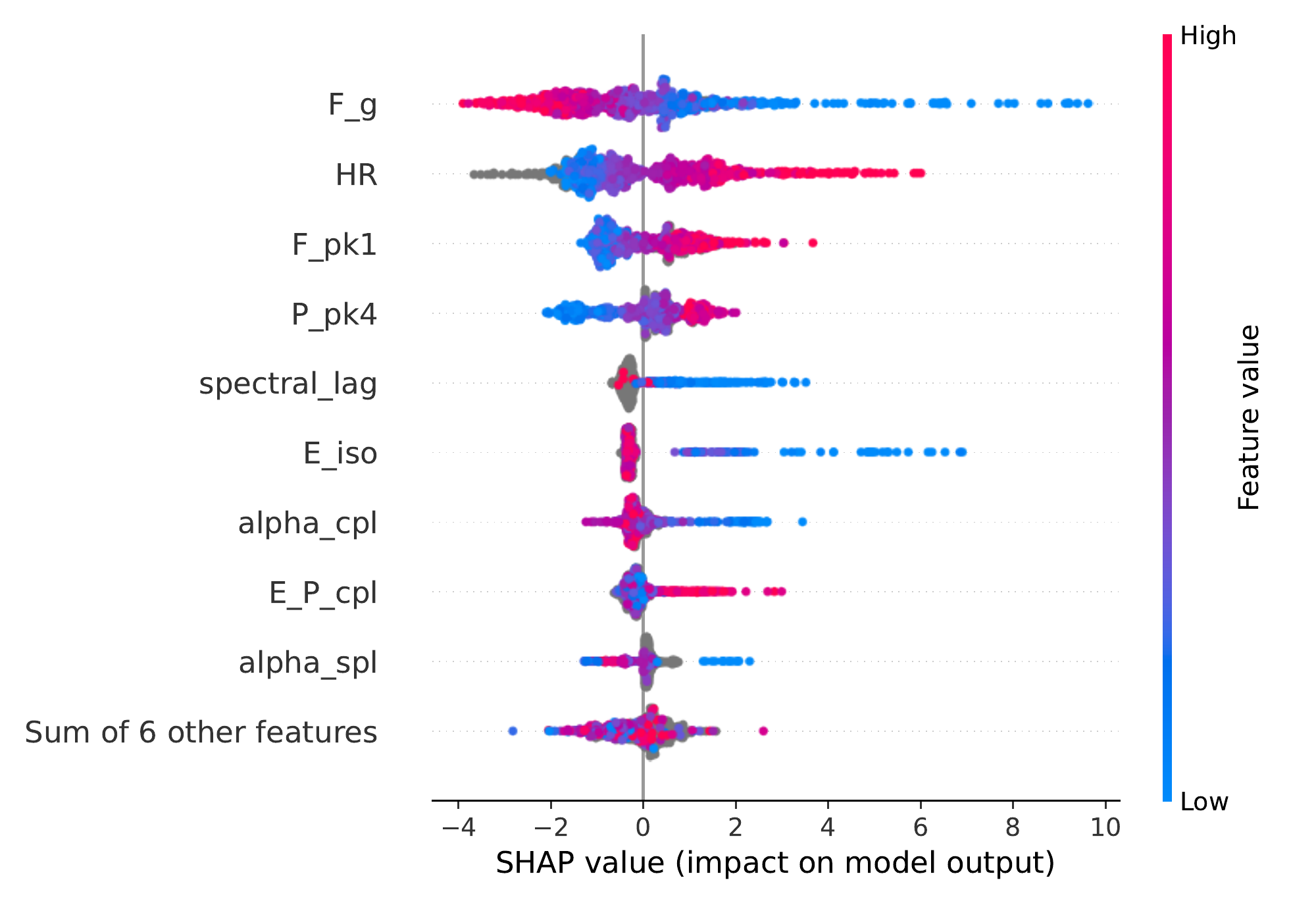}
    \label{fig:prompt_no_t90_fi2}}
\caption{Examples of confusion matrices and SHAP feature importance values of the prompt emission subgroup without $T_{90}$.}
\label{fig:prompt_no_t90}
\end{figure*}

However, when we remove $T_{90}$ from the prompt emission subgroup and carry out the same analysis, while we get a lower $F_1$ score of 0.581 on the test set as expected, but we also get a higher $F_1$ score of 0.833 on the intermingled GRBs. The $F_1$ score for intermediate GRBs is at a similar value of 0.888. The corresponding confusion matrices and feature importance are shown in Figure \ref{fig:prompt_no_t90}. This shows that $T_{90}$ can be misleading to the machine learning model for intermingled GRBs, and multiple observational parameters are needed for more accurate classification of GRBs.

We also find the fluence \texttt{F\_g} and hardness ratio \texttt{HR} to be the most important feature after $T_{90}$. A lower fluence and a higher HR pull the predictions toward Type I. Since fluence is directly related to the duration, our results confirm the finding of other studies.

\begin{figure*}
\centering
\subfloat[Confusion matrix on all GRBs in the test set]{
    \includegraphics[width=0.48\textwidth]{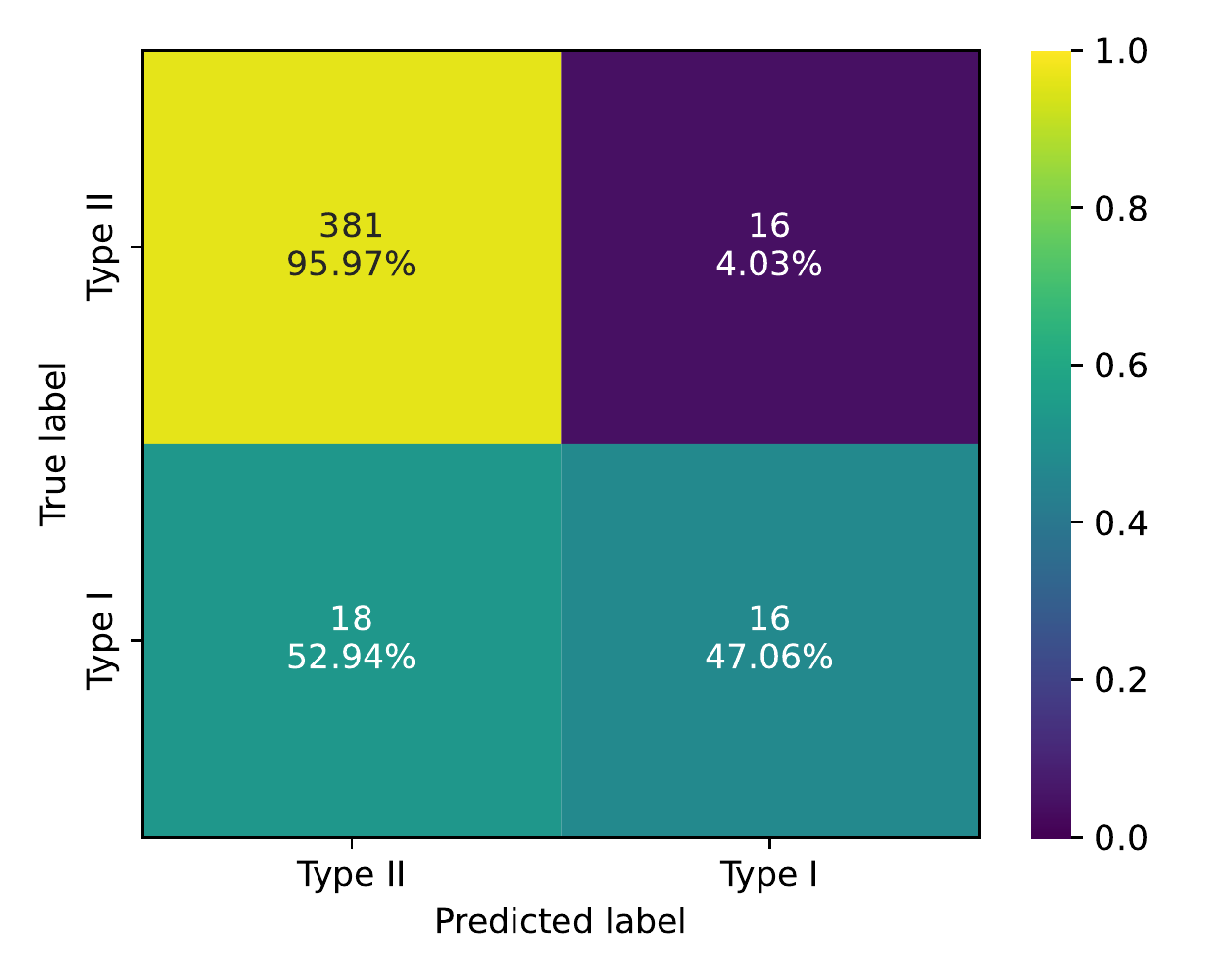}
    \label{fig:prompt_no_t90_fg_hr_cm}}
\subfloat[Confusion matrix on intermingled GRBs in the test set]{
    \includegraphics[width=0.48\textwidth]{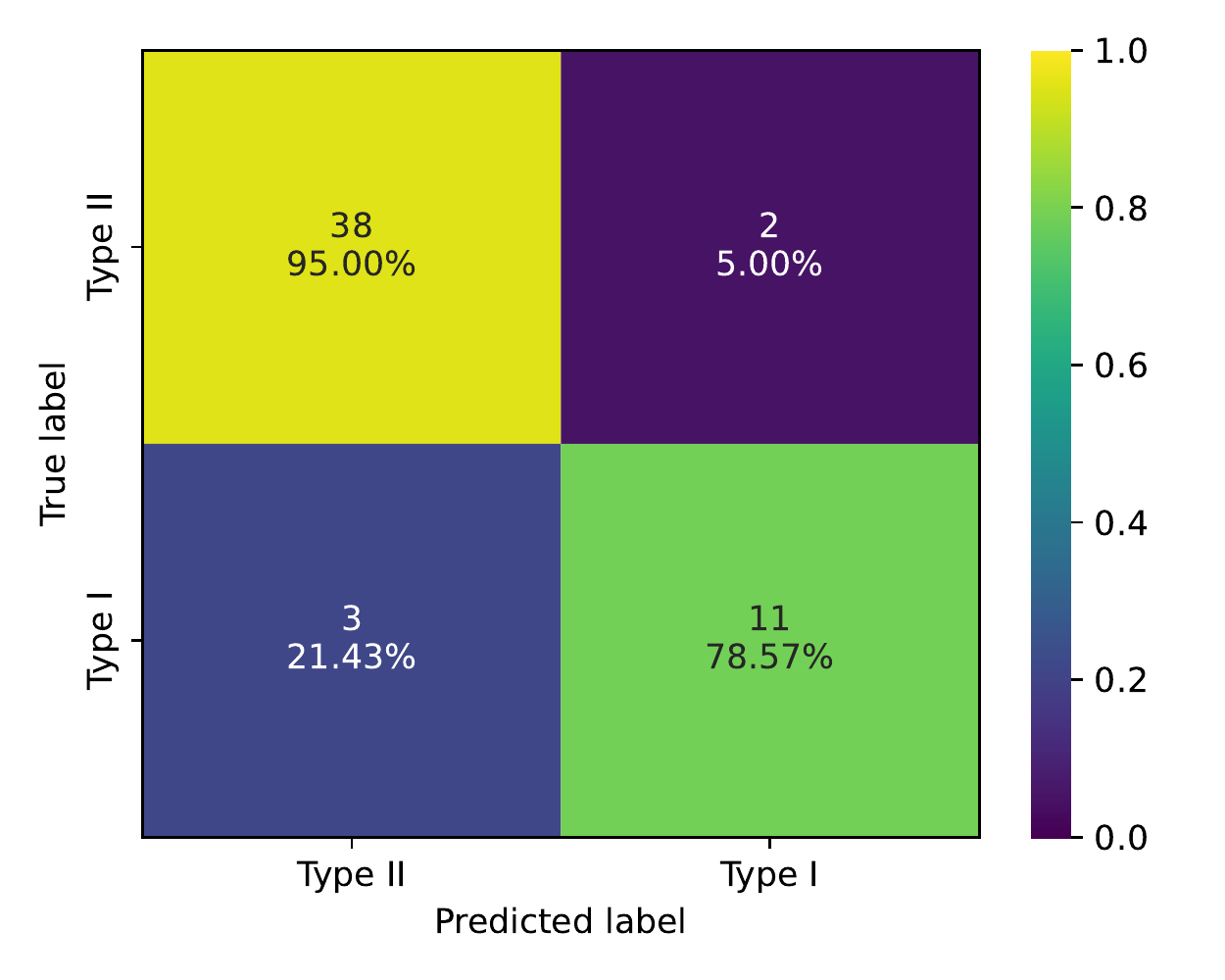}
    \label{fig:prompt_no_t90_fg_hr_inter_cm}}\\
\subfloat[Confusion matrix on intermediate GRBs in the test set]{
    \includegraphics[width=0.48\textwidth]{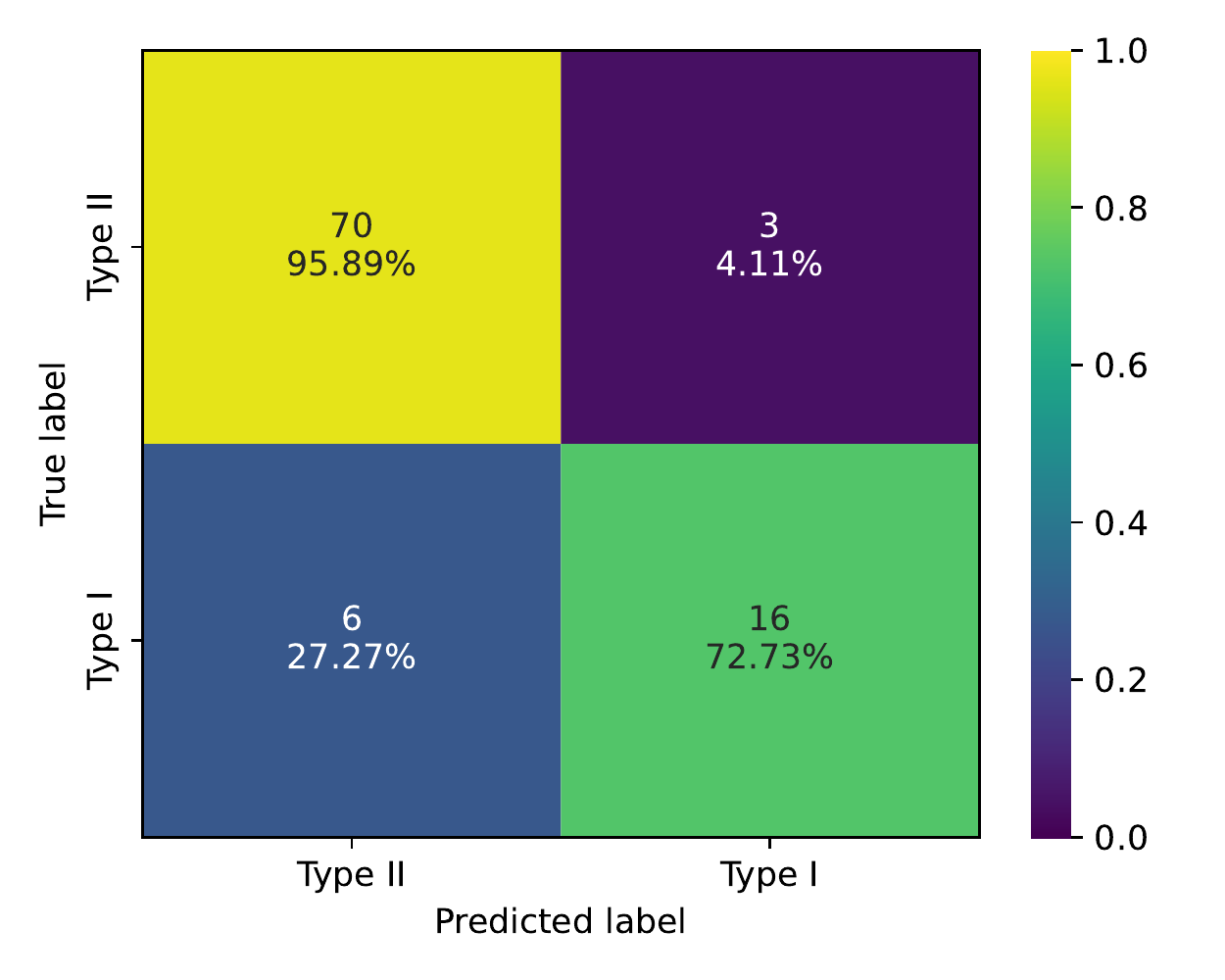}
    \label{fig:prompt_no_t90_fg_hr_mediate_cm}}
\subfloat[Average SHAP values of each feature on the training set]{
    \includegraphics[width=0.48\textwidth]{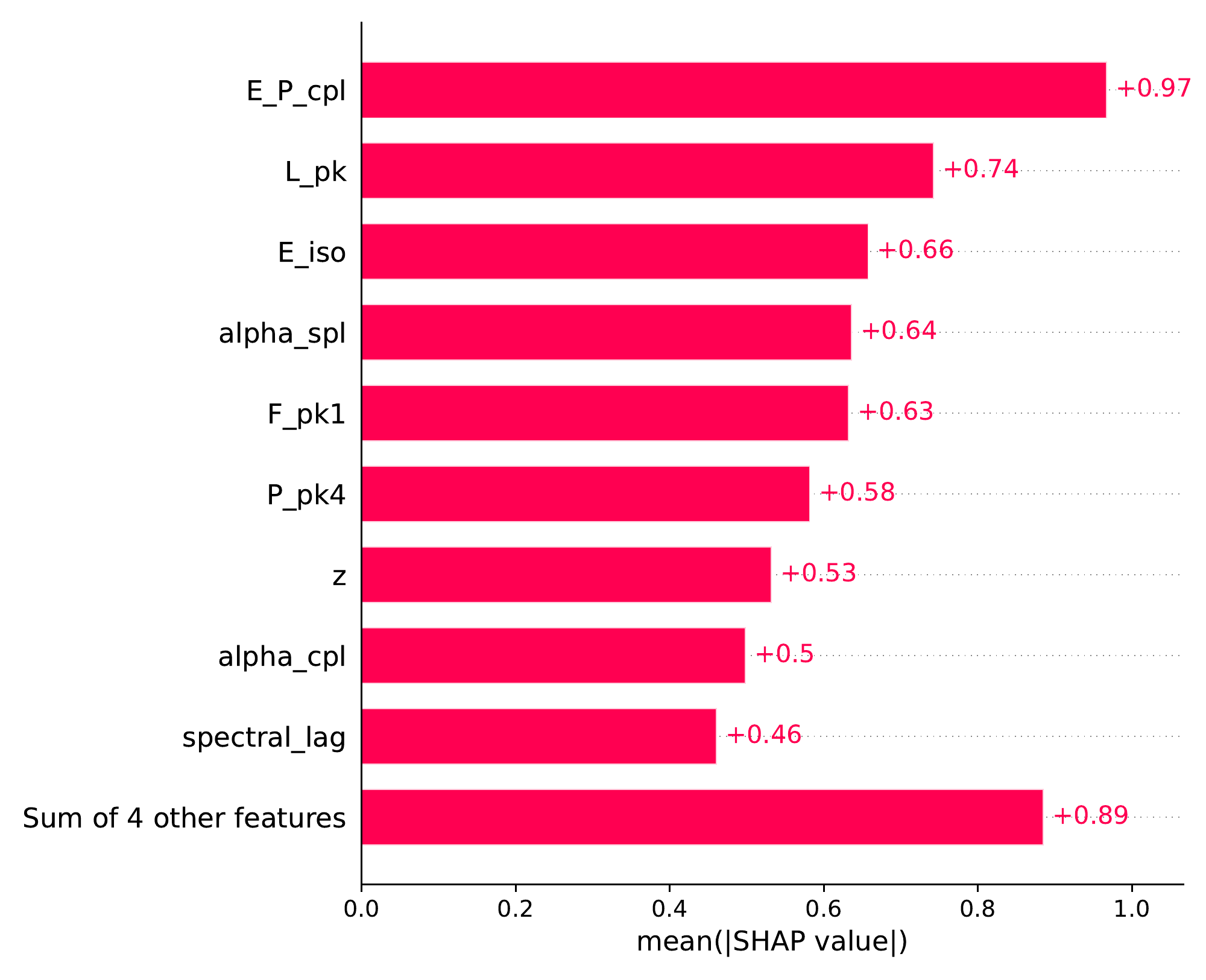}
    \label{fig:prompt_no_t90_fg_hr_fi}}\\
\subfloat[SHAP value beeswarm plot on the training set]{
    \includegraphics[width=0.48\textwidth]{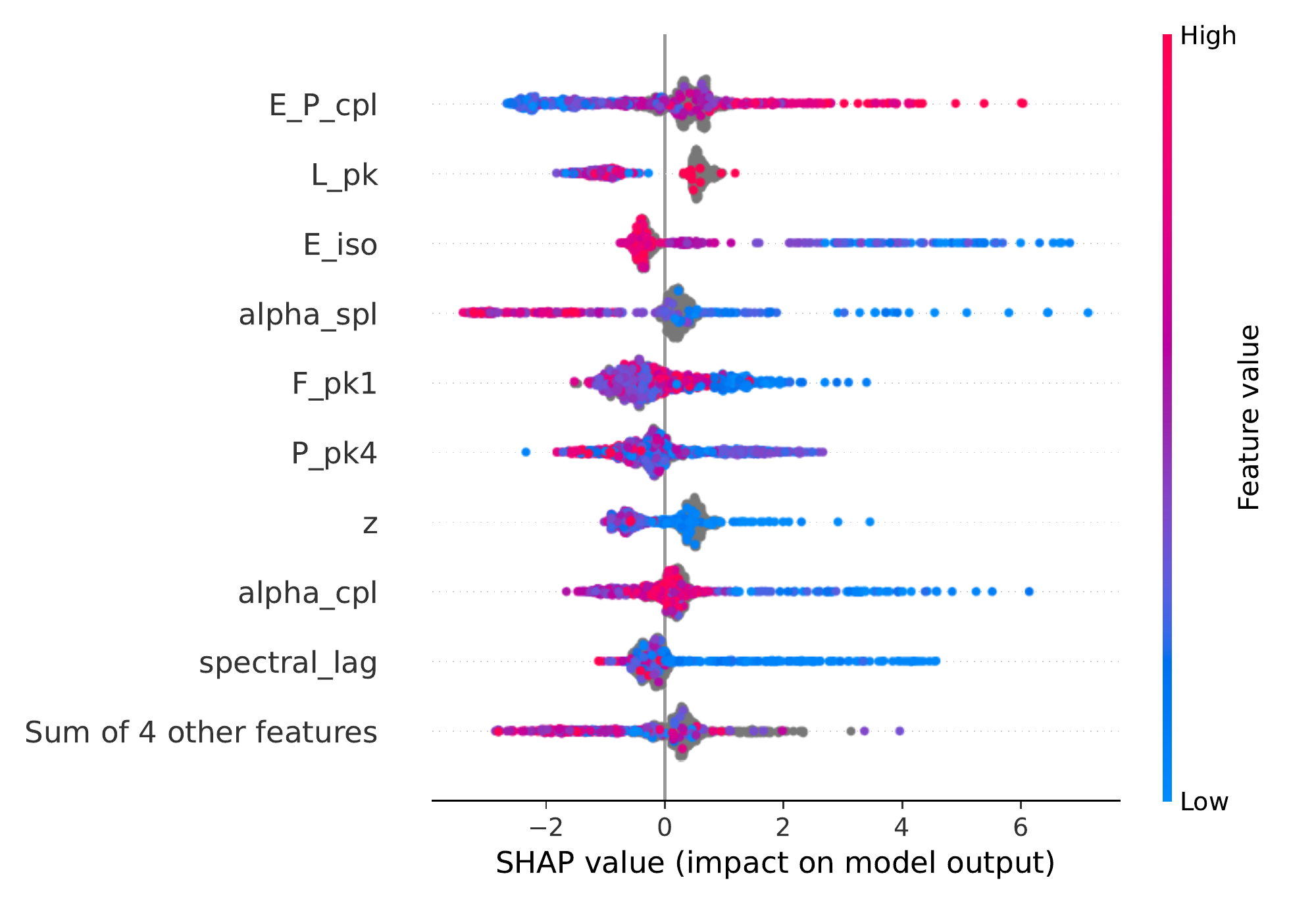}
    \label{fig:prompt_no_t90_fg_hr_fi2}}
\caption{Examples of confusion matrices and SHAP feature importance values of the prompt emission subgroup without $T_{90}$, \texttt{F\_g} or \texttt{HR}.}
\label{fig:prompt_no_t90_fg_hr}
\end{figure*}

In order to measure the importance of other features, we further exclude fluence and hardness ratio from our feature group, and carry out the same machine learning analysis. We obtain $F_1$ score of 0.485 on the test set, 0.815 on the intermingled GRBs and 0.780 on the intermediate GRBs. The corresponding confusion matrices and feature importance are shown in Figure \ref{fig:prompt_no_t90_fg_hr}. While the general $F_1$ score drops again, the $F_1$ scores for intermingled and intermediate samples remain high. The most important features are again related to the spectral shape, such as \texttt{E\_p\_cpl}, \texttt{E\_iso}, \texttt{alpha\_spl} and \texttt{alpha\_cpl}. The flux-related feature of \texttt{L\_pk}, \texttt{F\_pk1} and \texttt{P\_pk4} are also important, as well as redshift and spectral lag. Generally, a harder spectrum, a lower flux, a shorter spectral lag and a lower redshift pull the predictions toward Type I.

\subsection{Afterglow}
\label{subsec:afterglow}

\citet{gehrels2008CorrelationsPromptAfterglow,nysewander2009COMPARISONAFTERGLOWSSHORT,davanzo2012CompleteSampleBright,margutti2013PromptafterglowConnectionGammaray} pointed out that afterglows of Type I GRBs mostly have lower X-ray luminosities and energies. The X-ray luminosities and energies of Type I GRBs also decay faster. There are also correlations among afterglow X-ray energy, X-ray afterglow luminosity, prompt emission isotropic energy $E_{iso}$, peak luminosity $L_p$ and peak energy $E_p$. Combined with the findings mentioned in Section \ref{subsec:prompt_emission}, X-ray afterglow luminosity can also be employed for GRB classification.

\citet{kann2011AFTERGLOWSSWIFTERA} found that similar to X-ray, optical afterglows of Type I GRBs are significantly fainter than that of Type II GRBs, and similar afterglow-prompt emission correlations also exist in the optical band.

\begin{figure*}
\centering
\subfloat[Confusion matrix on all GRBs in the test set]{%
    \includegraphics[width=0.48\textwidth]{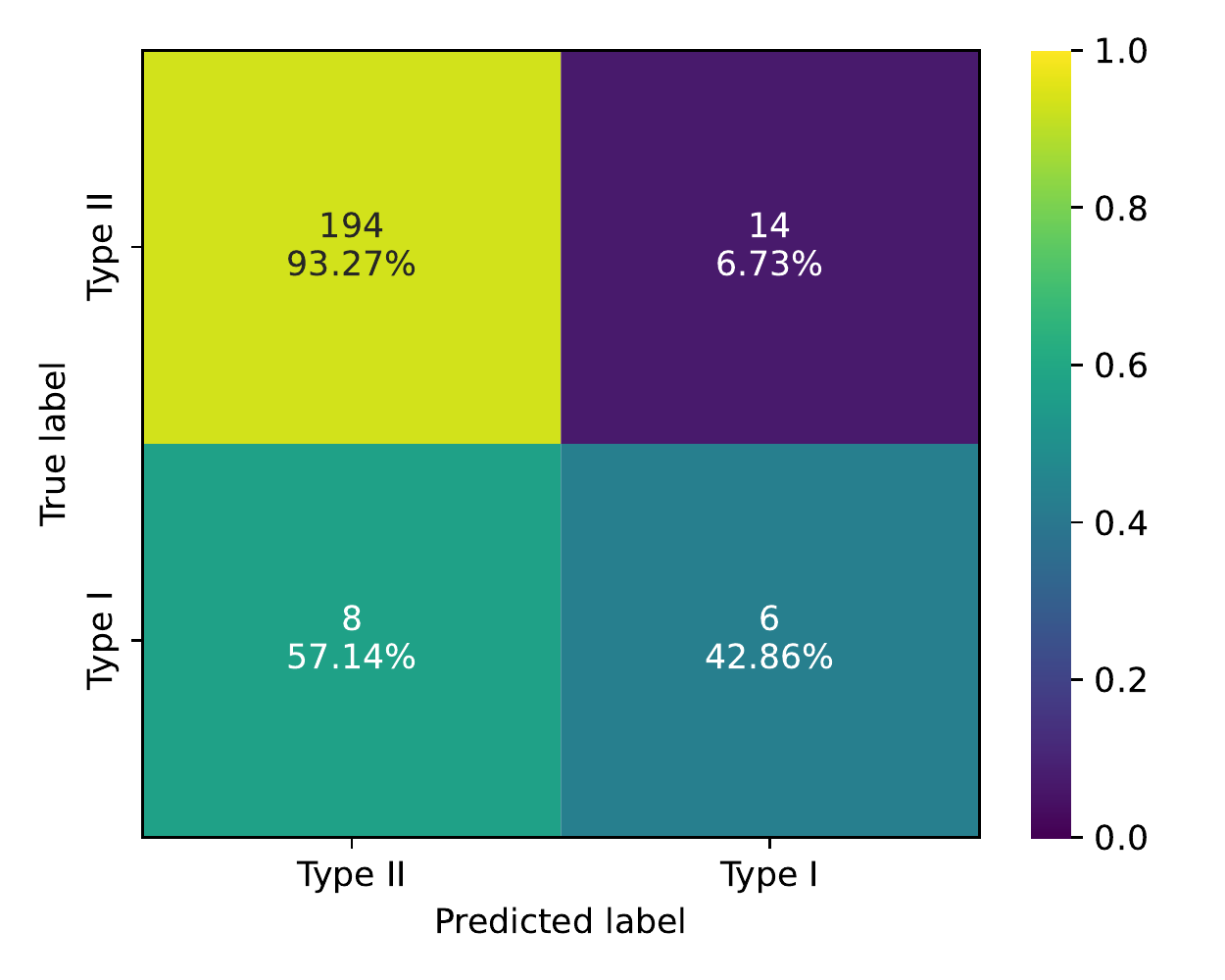}%
    \label{fig:afterglow_cm}}
\subfloat[Confusion matrix on intermingled GRBs in the test set]{%
    \includegraphics[width=0.48\textwidth]{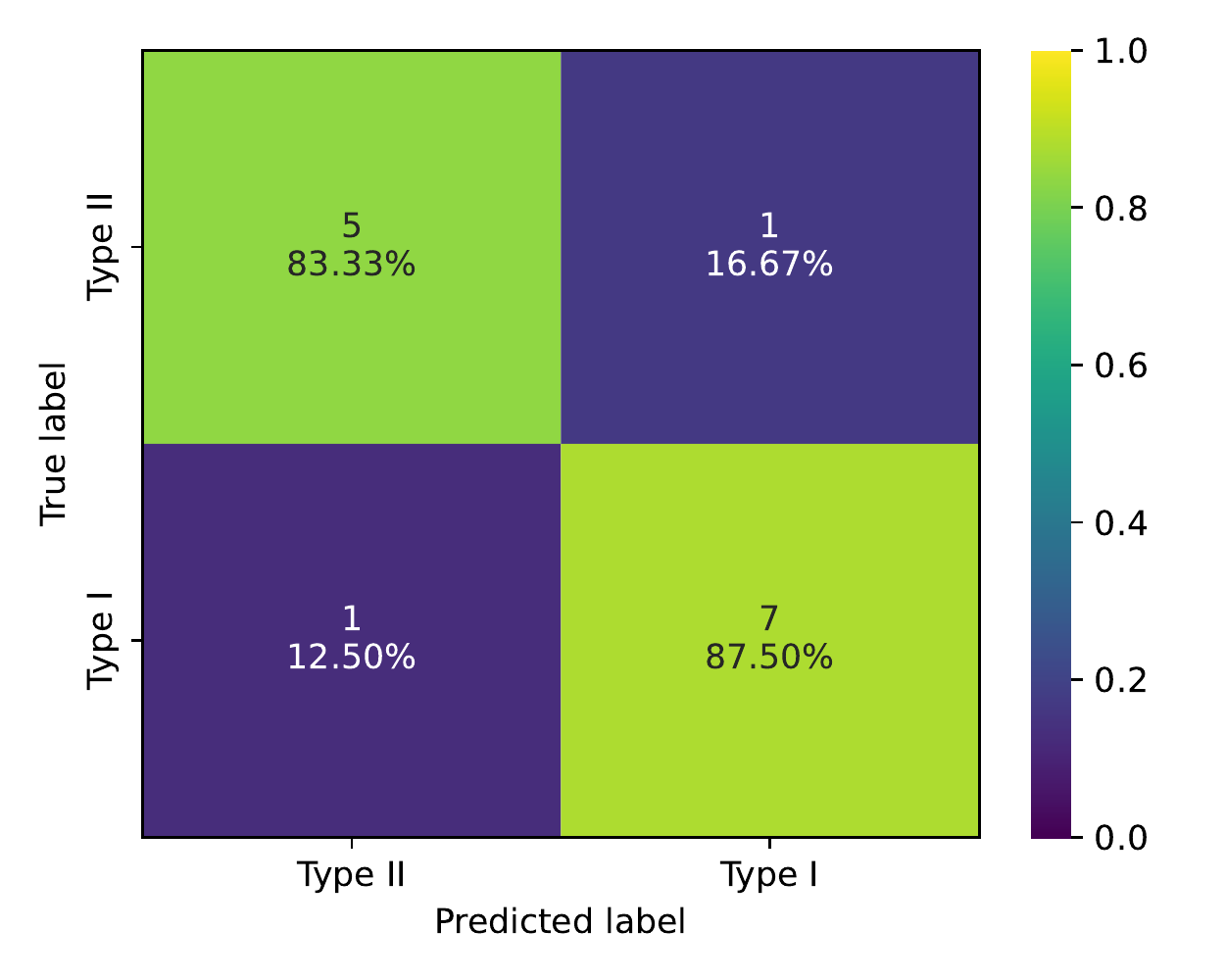}%
    \label{fig:afterglow_inter_cm}}\\
\subfloat[Confusion matrix on intermediate GRBs in the test set]{%
    \includegraphics[width=0.48\textwidth]{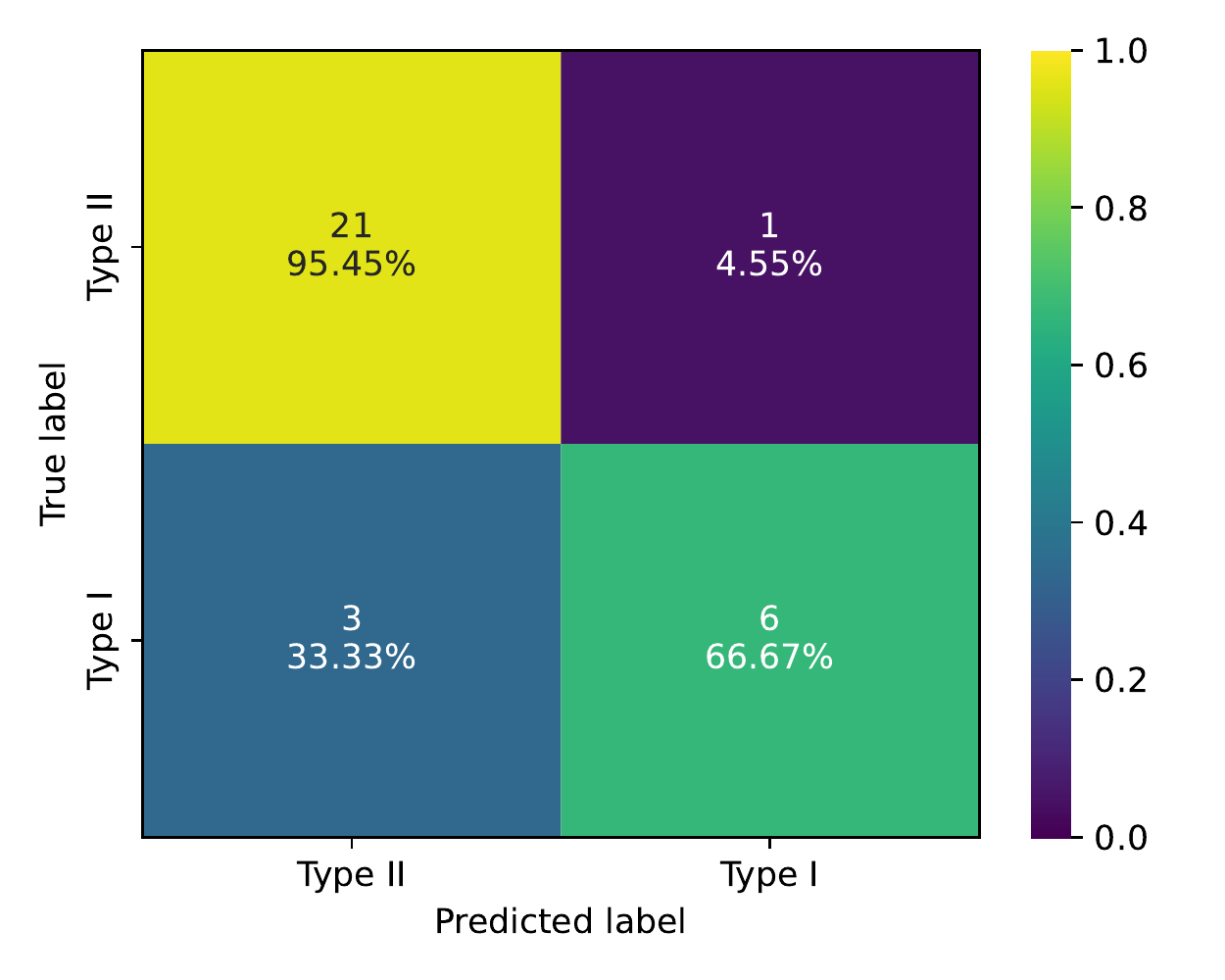}%
    \label{fig:afterglow_mediate_cm}}
\subfloat[Average SHAP values of each feature on the training set]{%
    \includegraphics[width=0.48\textwidth]{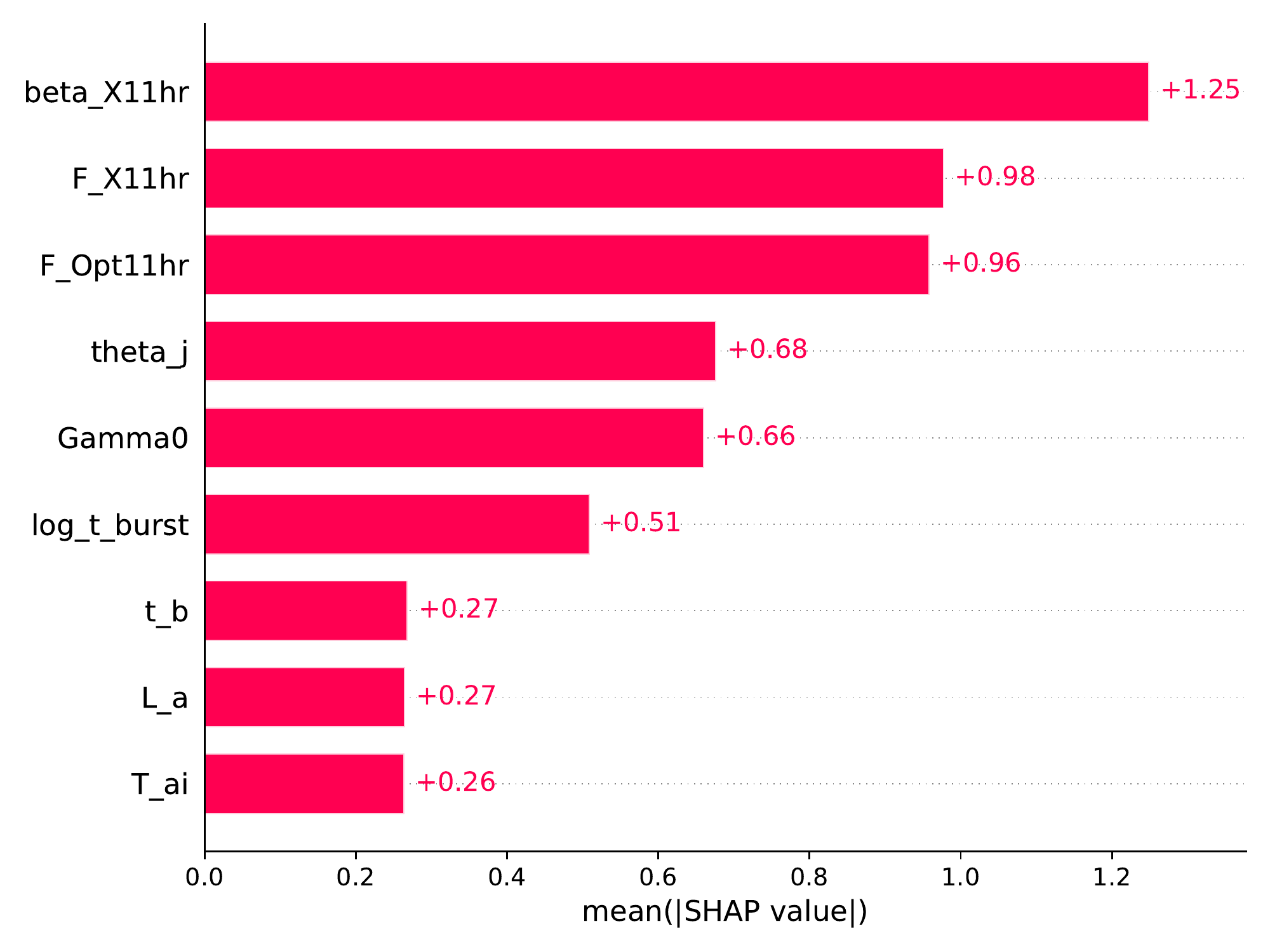}%
    \label{fig:afterglow_fi}}\\
\subfloat[SHAP value beeswarm plot on the training set]{%
    \includegraphics[width=0.48\textwidth]{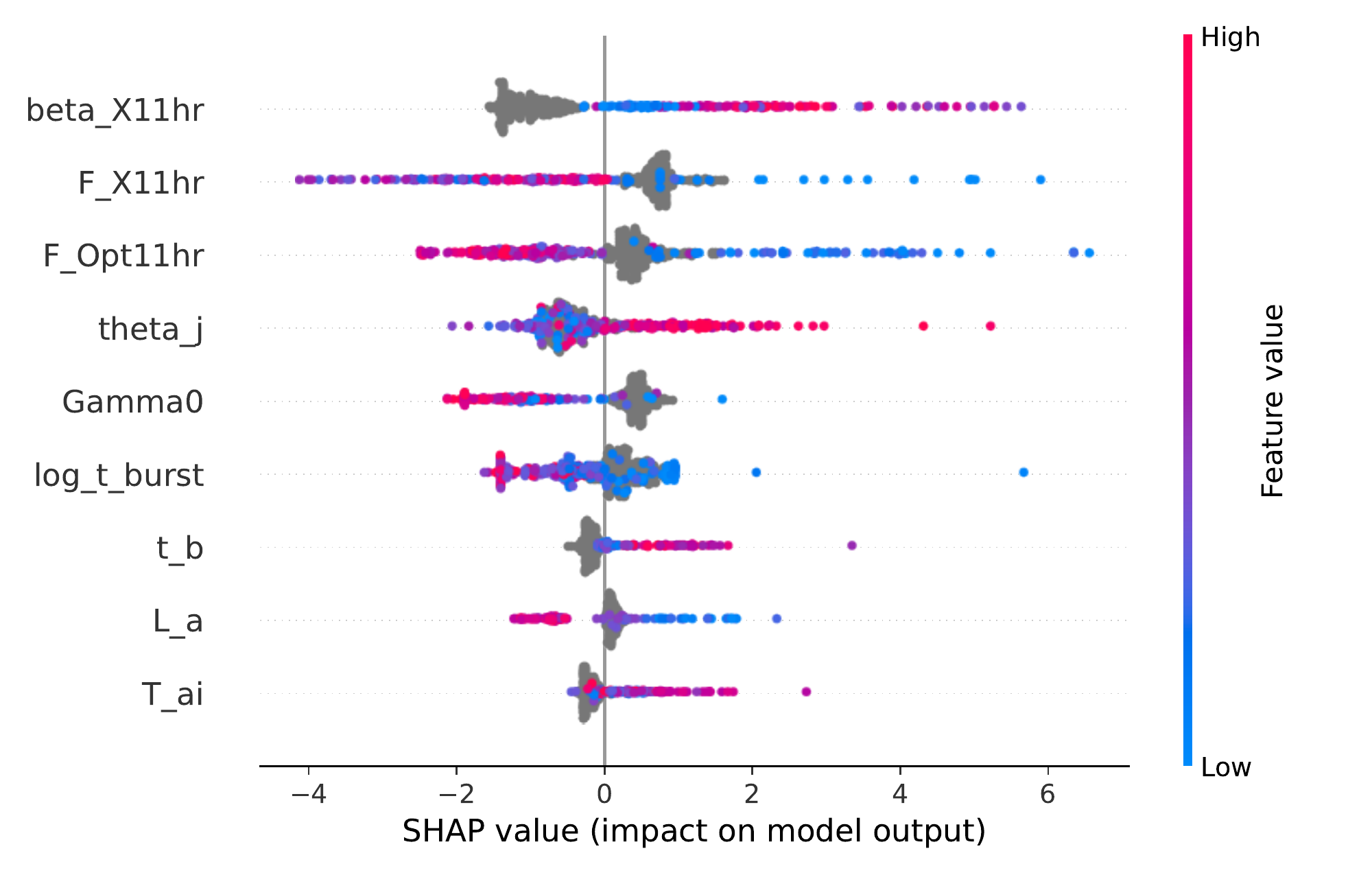}%
    \label{fig:afterglow_fi2}}
\caption{Examples of confusion matrices and SHAP feature importance values of the afterglow subgroup.}
\label{fig:afterglow}
\end{figure*}

With the afterglow subgroup, we are able to obtain $F_1$ score of 0.353 on the test set, 0.857 on the intermingled GRBs and 0.75 on the intermediate GRBs. The corresponding confusion matrices and feature importance are shown in Figure \ref{fig:afterglow}. We found the most important feature to be 11-hour beta index in X-rays. The 11-hour fluxes in X-ray and optical bands are also important. A higher beta index and lower X-ray and optical fluxes pull the predictions toward Type I, consistent with other studies. In general, we find that afterglow features perform poorly in GRB classification.

\subsection{Host galaxy}
\label{subsec:host_galaxy}

The different progenitors of Type I and II GRBs also have a substantial correlation with the properties of their host galaxies. The short lifetime of Type II GRB progenitors \citep{woosley2002EvolutionExplosionMassive} makes their event rate to generally follow the star formation rate (SFR) of the host galaxies, and Type II GRB host galaxies generally have higher SFR. \citep{bloom2002ObservedOffsetDistribution,chary2007SpitzerObservationsGamma,savaglio2009GALAXYPOPULATIONHOSTING,levesque2010HOSTGALAXIESGAMMARAYa,robertson2011CONNECTINGGAMMARAY,levesque2014HostGalaxiesLongDuration,wei2014CosmologicalTestsUsing,trenti2015LUMINOSITYLARMASS,cucchiara2015UNVEILINGSECRETSMETALLICITY,lan2022StellarmassFunctionLong}. The redshift distribution of Type I GRBs are found to be delayed with respect to the star formation history, and thus host galaxies of Type I GRBs generally have lower SFR respectively \citep{piran1992ImplicationsComptonGRO,nakar2006LocalRateProgenitor,zheng2007DeducingLifetimeShort,virgili2011AREALLSHORTHARD,wanderman2015RateLuminosityFunction,luo2022GlobalTestJet}.

Type II GRB hosts also have low metallicity, which is required to form high-mass progenitors. \citep{fynbo2003LyaEmissionGammaray,prochaska2004HostGalaxyGRB,fruchter2006LongGrayBursts,levesque2010HOSTGALAXIESGAMMARAY,kocevski2011ORIGINMASSMETALLICITY,mannucci2011MetallicityLongGRB,campisi2011MetallicityPropertiesSimulated,graham2017RELATIVERATELGRB,lesniewska2022InterstellarMediumEnvironment}. Type I GRB hosts, on the other hand, are found to have higher metallicity \citep[e.g.][]{berger2014ShortDurationGammaRayBursts}.

Type II GRBs usually occur in regions with active star formation and are, therefore, closer to the center of the galaxy and in brighter regions. Type I GRBs, however, have larger offsets from the galactic center as the evolution of compact binary mergers require supernova events that ``kick off" the binary system away from the location where they are formed \citep{bloom2002ObservedOffsetDistribution,fruchter2006LongGrayBursts,fong2013DEMOGRAPHICSGALAXIESHOSTING,blanchard2016OFFSETHOSTLIGHT,wang2018PossibleCorrelationsGammaray,li2020ComparativeStudyLong,oconnor2022DeepSurveyShort,fong2022ShortGRBHost}.

\begin{figure*}
\centering
\subfloat[Confusion matrix on all GRBs in the test set]{
    \includegraphics[width=0.48\textwidth]{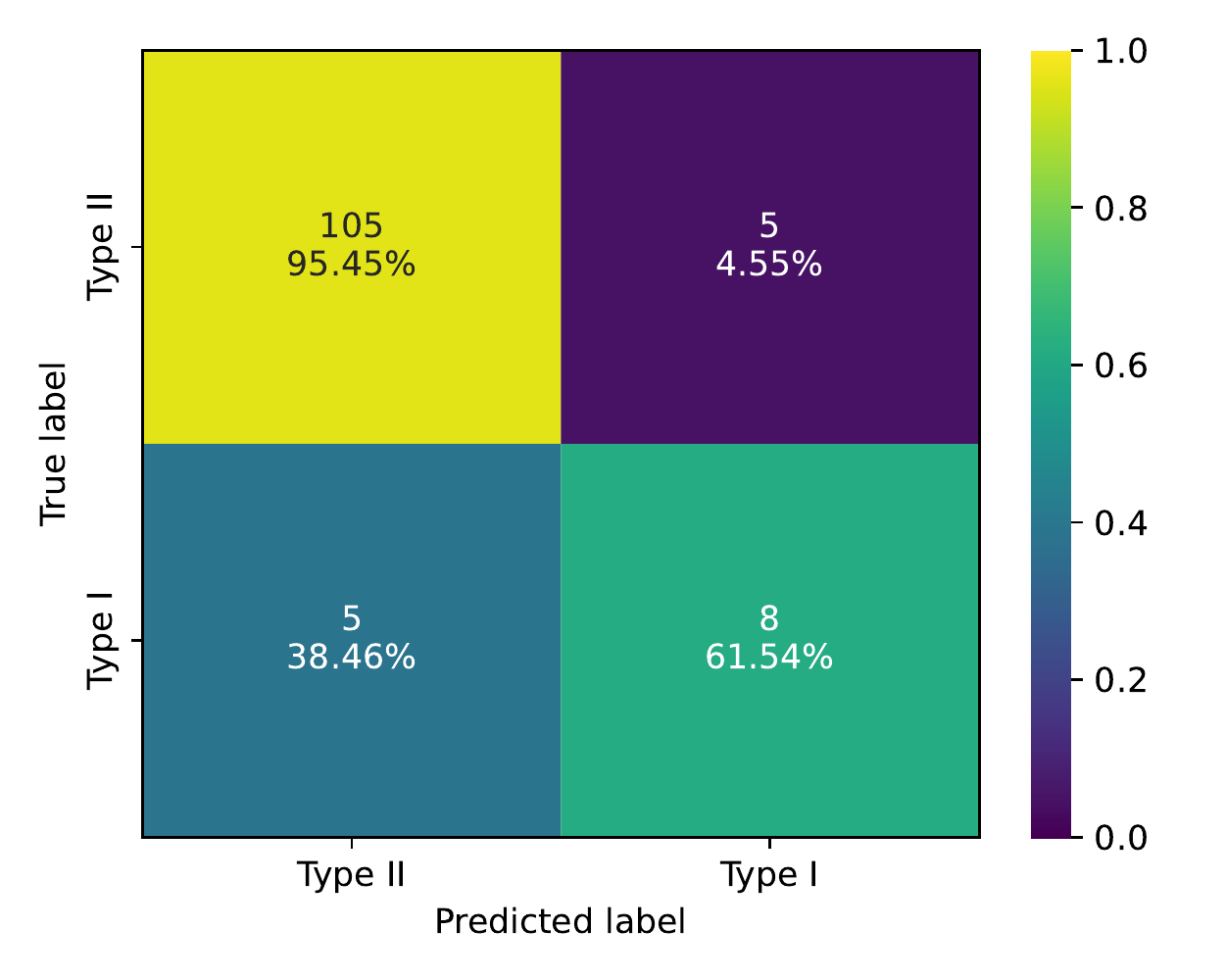}
    \label{fig:host_cm}}
\subfloat[Confusion matrix on intermingled GRBs in the test set]{
    \includegraphics[width=0.48\textwidth]{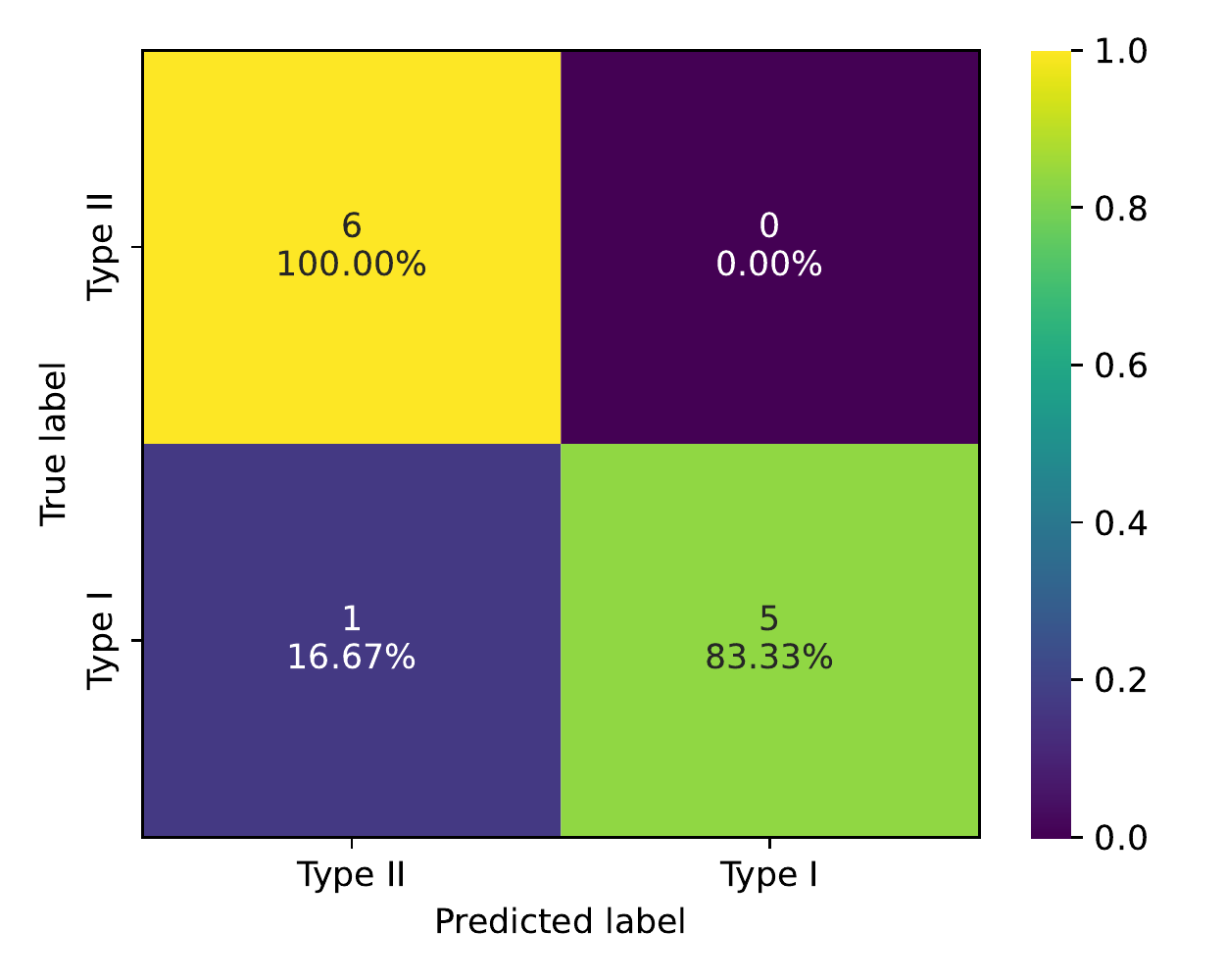}
    \label{fig:host_inter_cm}}\\
\subfloat[Confusion matrix on intermediate GRBs in the test set]{
    \includegraphics[width=0.48\textwidth]{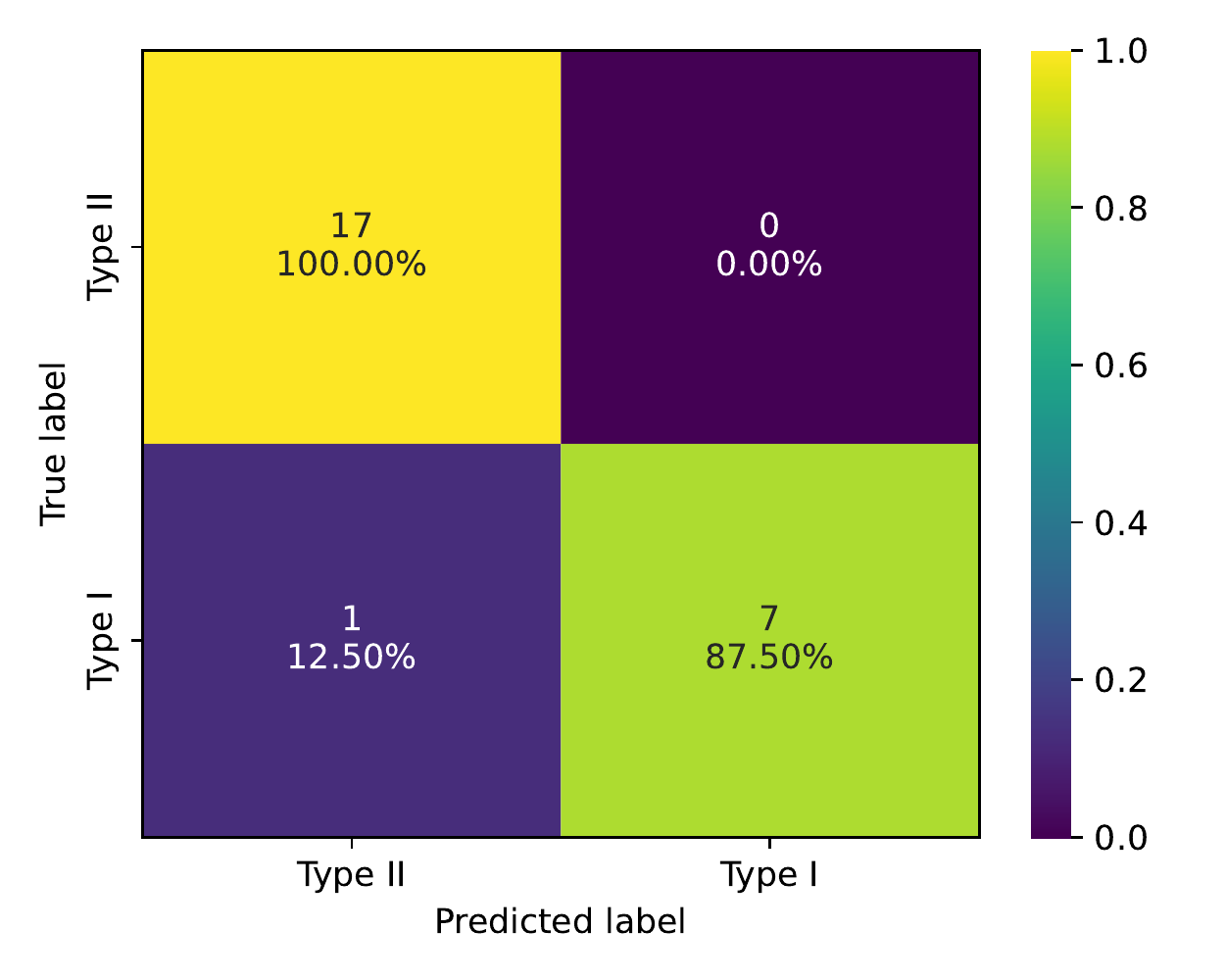}
    \label{fig:host_mediate_cm}}
\subfloat[Average SHAP values of each feature on the training set]{
    \includegraphics[width=0.48\textwidth]{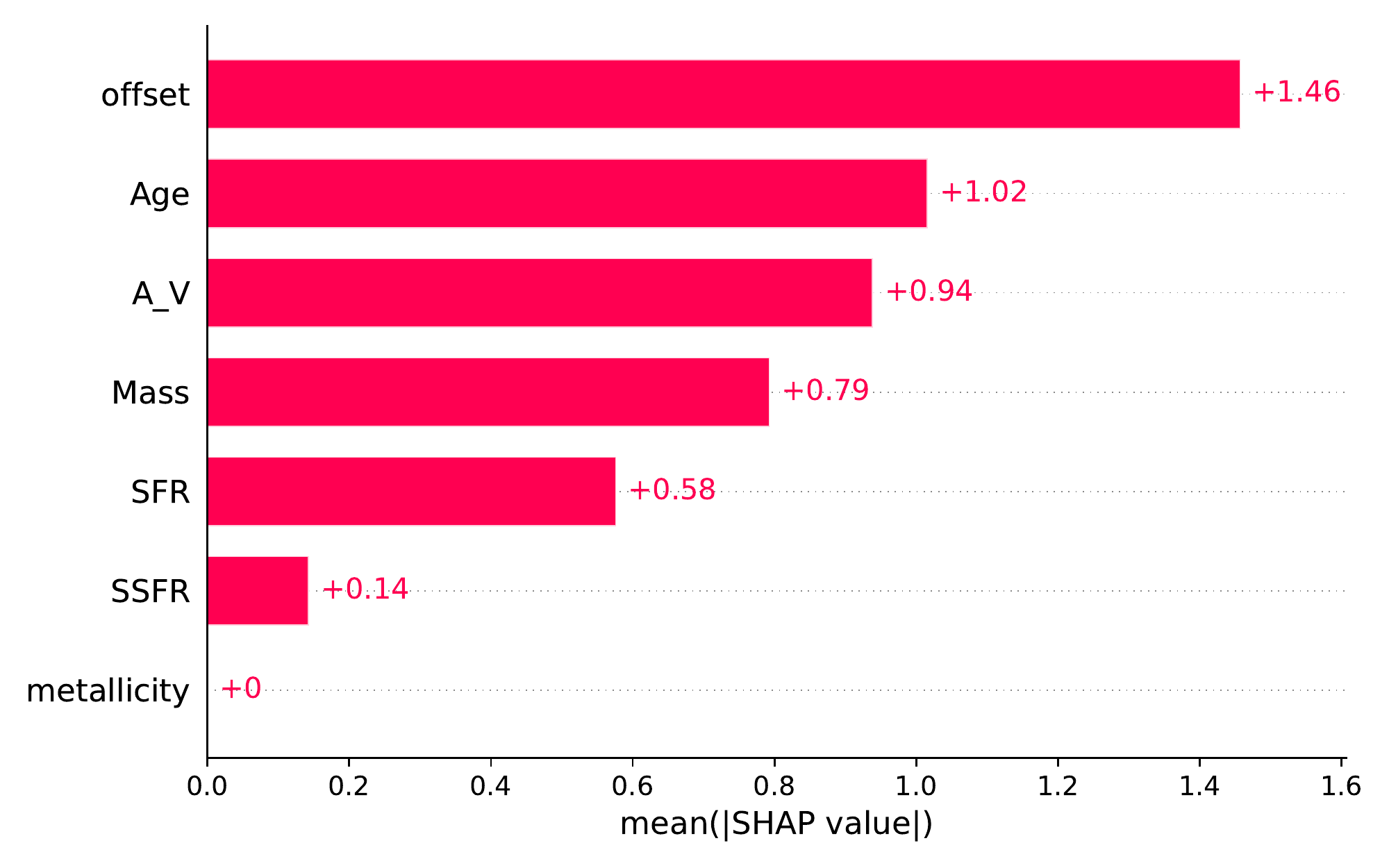}
    \label{fig:host_fi}}\\
\subfloat[SHAP value beeswarm plot on the training set]{
    \includegraphics[width=0.48\textwidth]{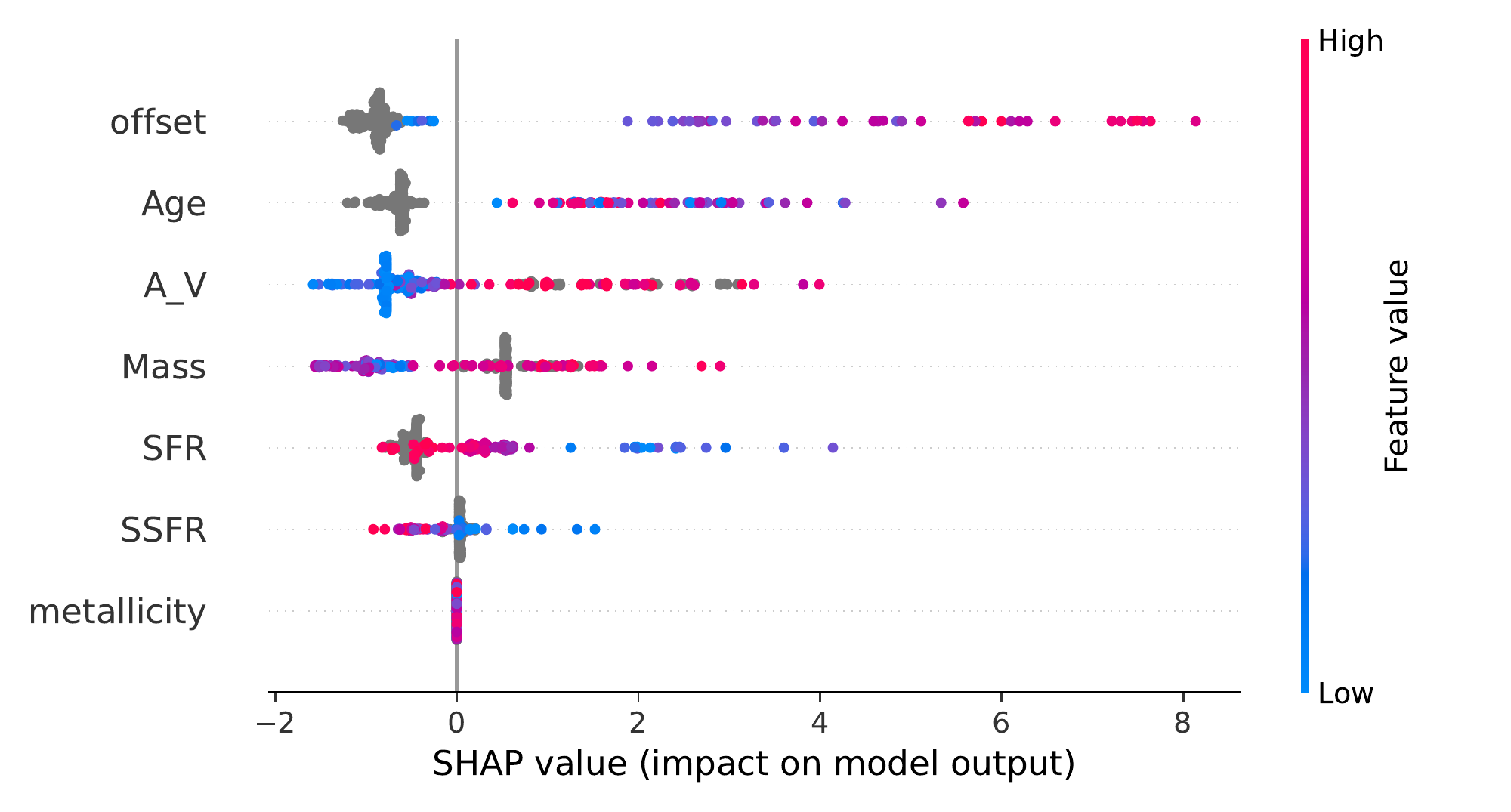}
    \label{fig:host_fi2}}
\caption{Examples of confusion matrices and SHAP feature importance values of the host galaxy subgroup.}
\label{fig:host}
\end{figure*}

With the host galaxy subgroup, we are able to obtain $F_1$ score of 0.615 on the test set, 0.909 on the intermingled GRBs and 0.933 on the intermediate GRBs. The corresponding confusion matrices and feature importance are shown in Figure \ref{fig:host}. We found the most important feature to be offset, with higher offset pull the predictions toward Type I.

\begin{figure*}
\centering
\subfloat[Confusion matrix on all GRBs in the test set]{
    \includegraphics[width=0.49\textwidth]{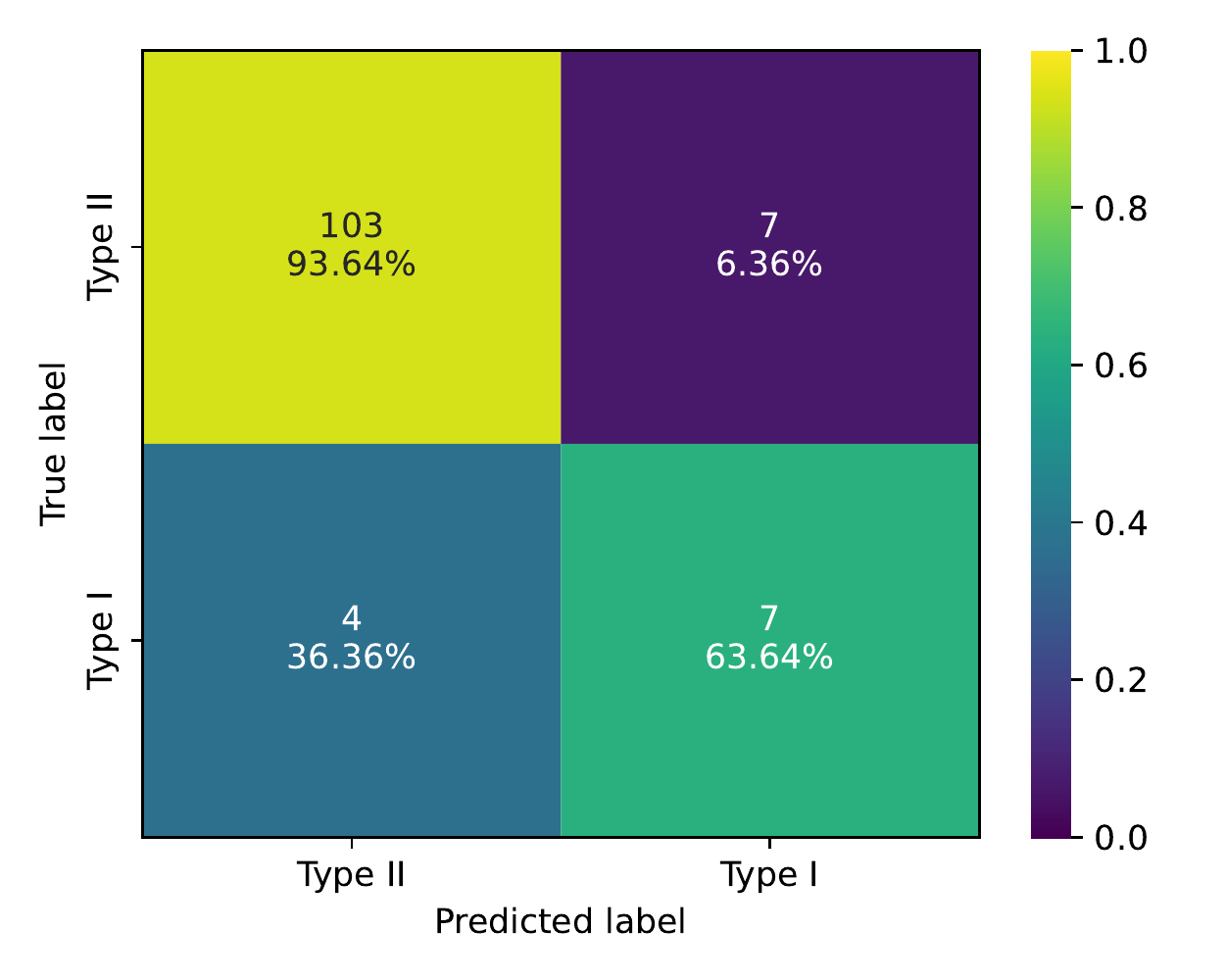}
    \label{fig:host_no_offset_cm}}
\subfloat[Confusion matrix on intermingled GRBs in the test set]{
    \includegraphics[width=0.49\textwidth]{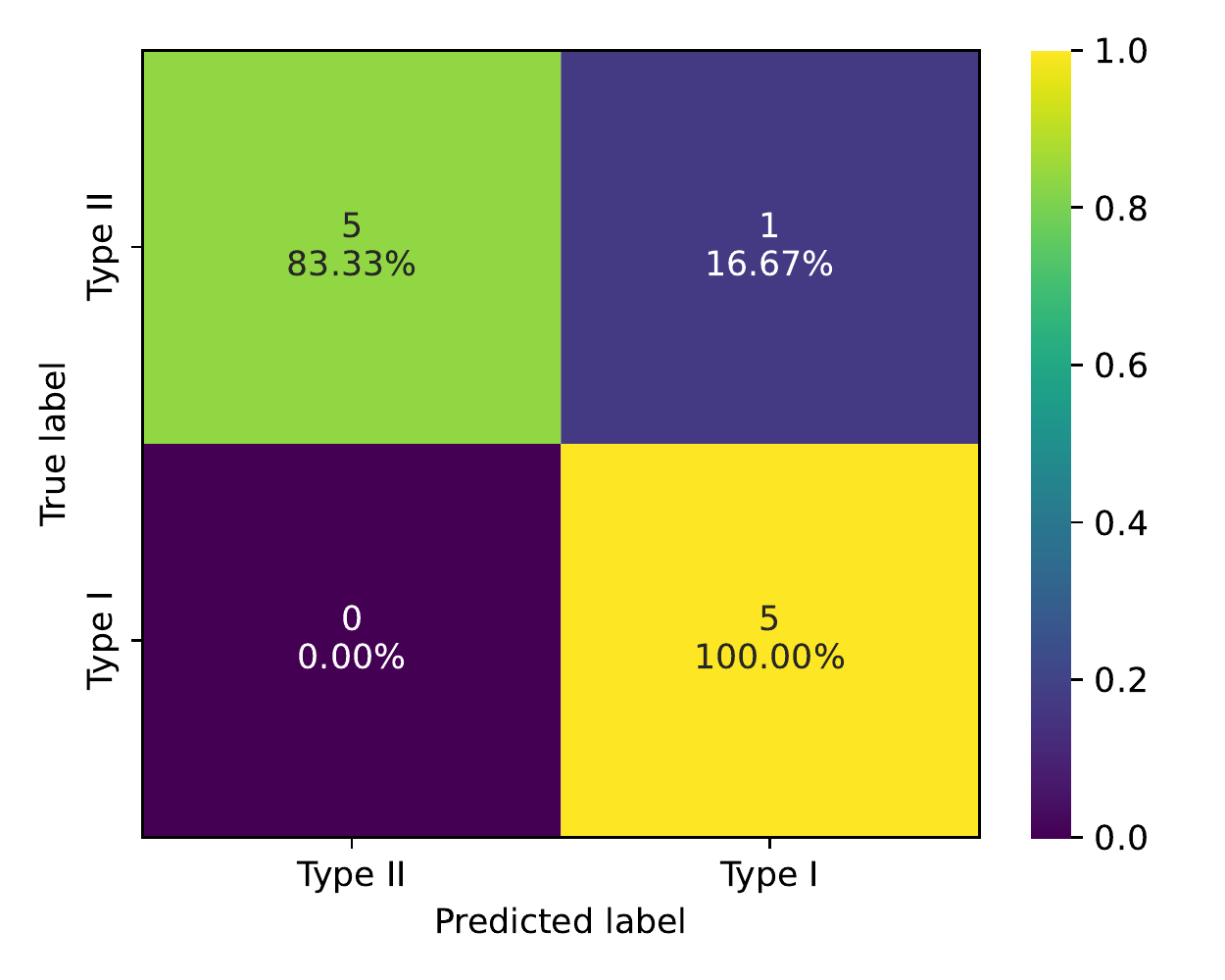}
    \label{fig:host_no_offset_inter_cm}}\\
\subfloat[Confusion matrix on intermediate GRBs in the test set]{
    \includegraphics[width=0.49\textwidth]{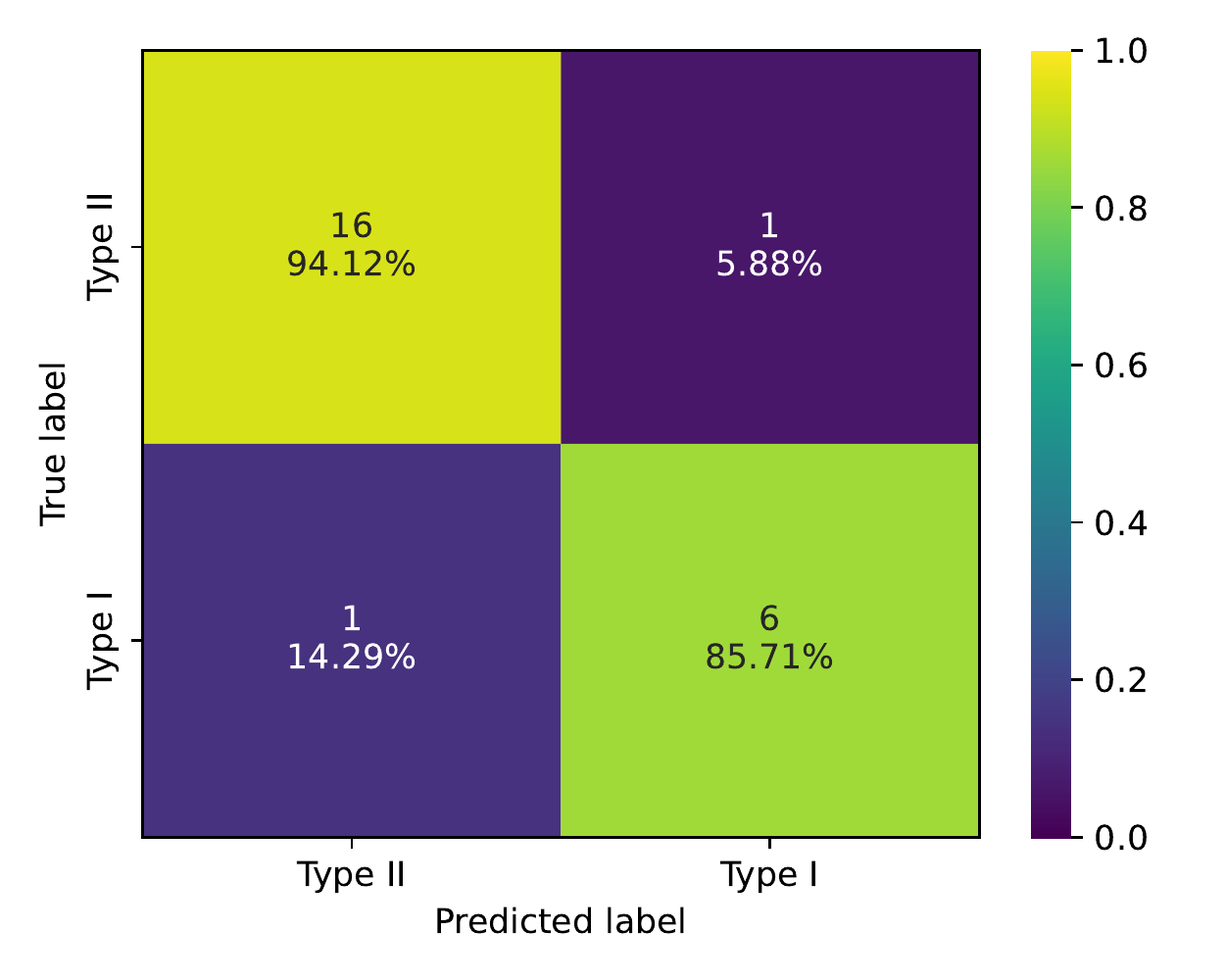}
    \label{fig:host_no_offset_mediate_cm}}
\subfloat[Average SHAP values of each feature on the training set]{
    \includegraphics[width=0.49\textwidth]{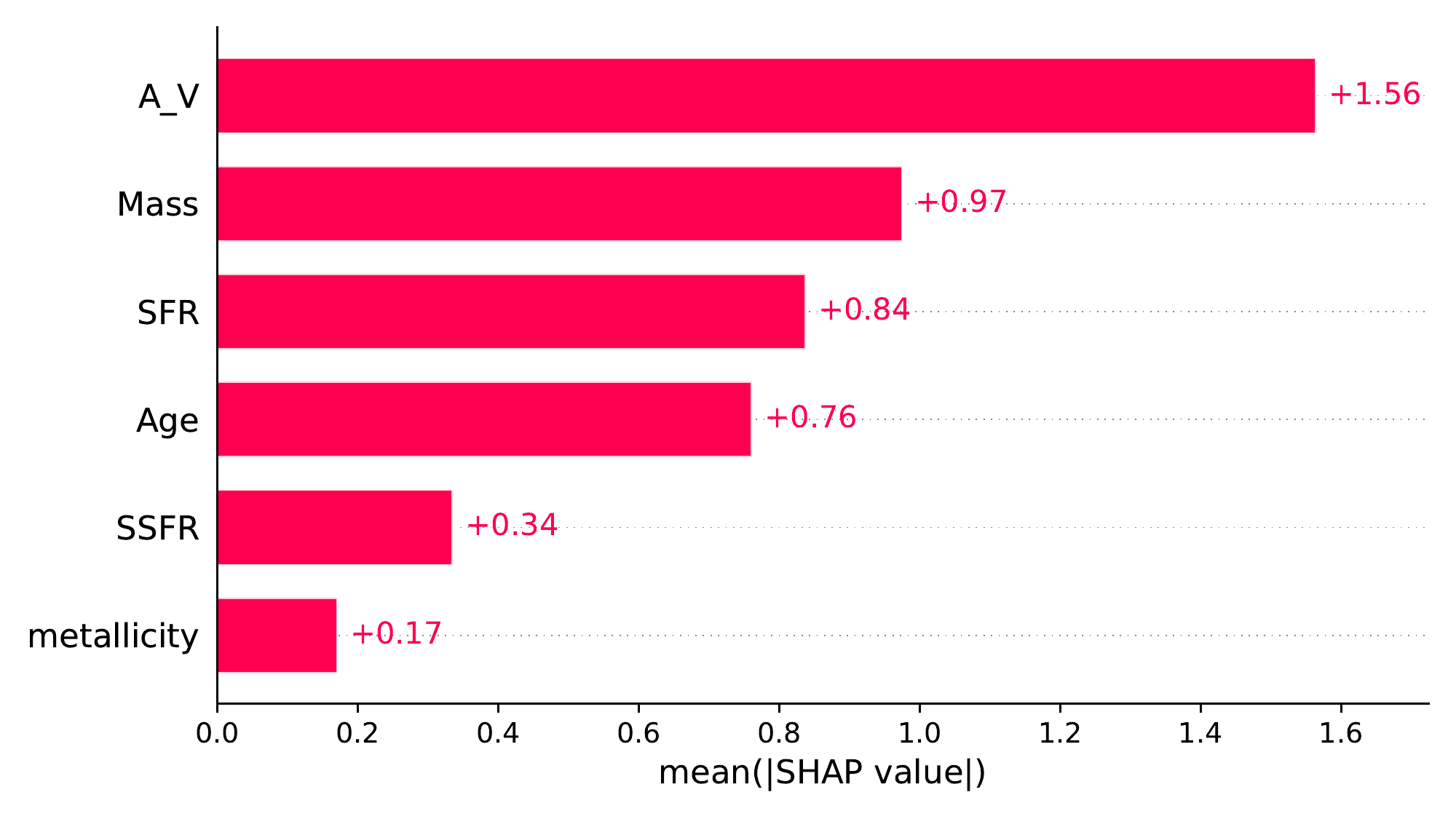}
    \label{fig:host_no_offset_fi}}\\
\subfloat[SHAP value beeswarm plot on the training set]{
    \includegraphics[width=0.49\textwidth]{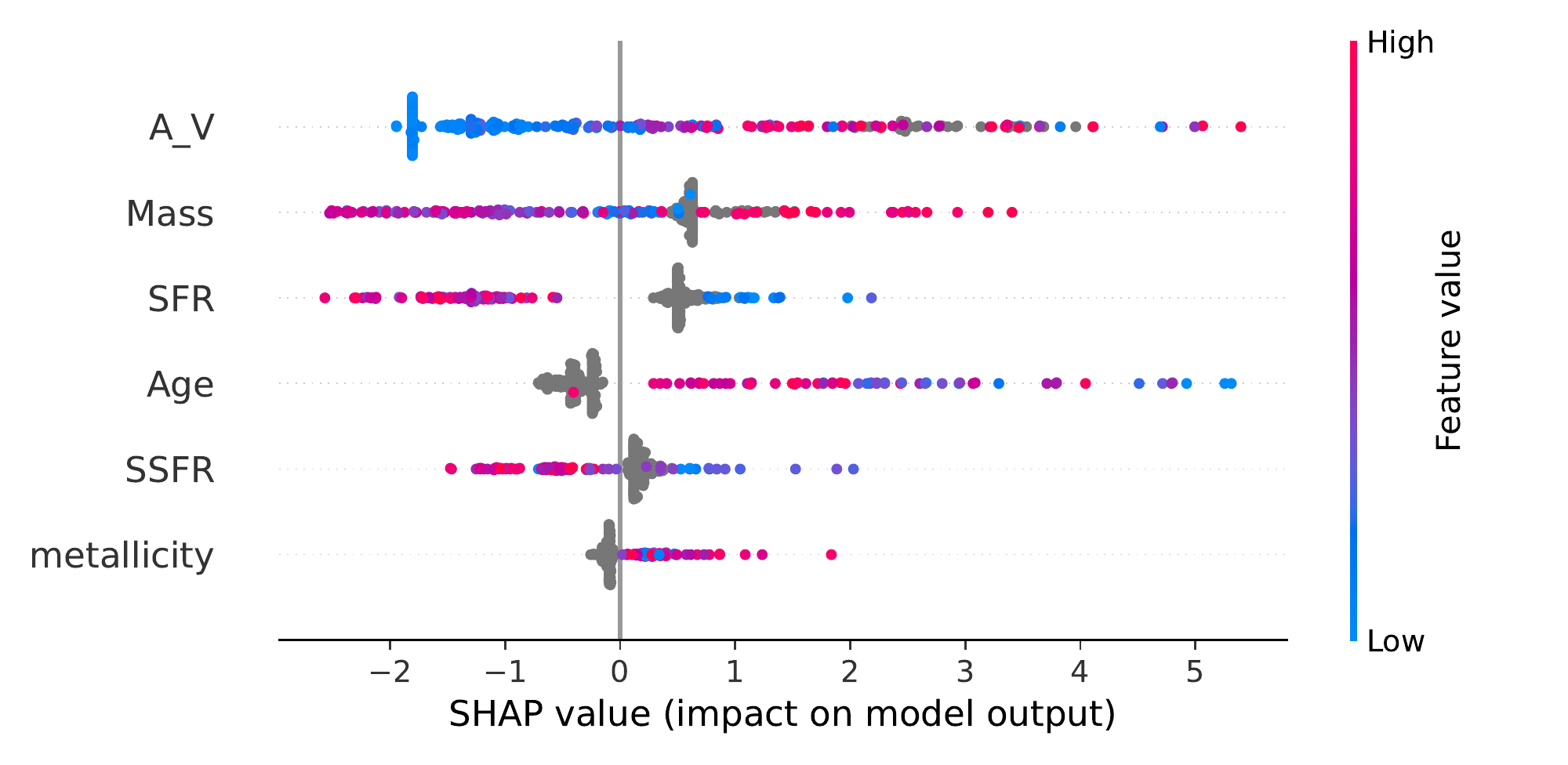}
    \label{fig:host_no_offset_fi2}}
\caption{Examples of confusion matrices and SHAP feature importance values of the host galaxy subgroup without \texttt{offset}.}
\label{fig:host_no_offset}
\end{figure*}

In order to find other important features, we also carry out the same analysis on the host galaxy subgroup without offset. We get $F_1$ score of 0.56 on the test set, 0.909 on the intermingled GRBs and 0.857 on the intermediate GRBs with the host galaxy subgroup without offset. The corresponding confusion matrices and feature importance are shown in Figure \ref{fig:host_no_offset}. \texttt{A\_V}, stellar mass and star formation rate (SFR) are fairly important. Stronger dust extinction, higher stellar mass and lower SFR pull the predictions toward Type I.


\subsection{All}
\label{subsec:all}

We also combine all the feature subgroups to form an "all" group. We then train and test our machine learning model with this group containing all the features. With all the features, we obtain a $F_1$ score of 0.8 on the test set, 0.649 on the intermingled GRBs and 0.870 on the intermediate GRBs.

\begin{figure*}
\centering
\subfloat[Confusion matrix on all GRBs in the test set]{
    \includegraphics[width=0.48\textwidth]{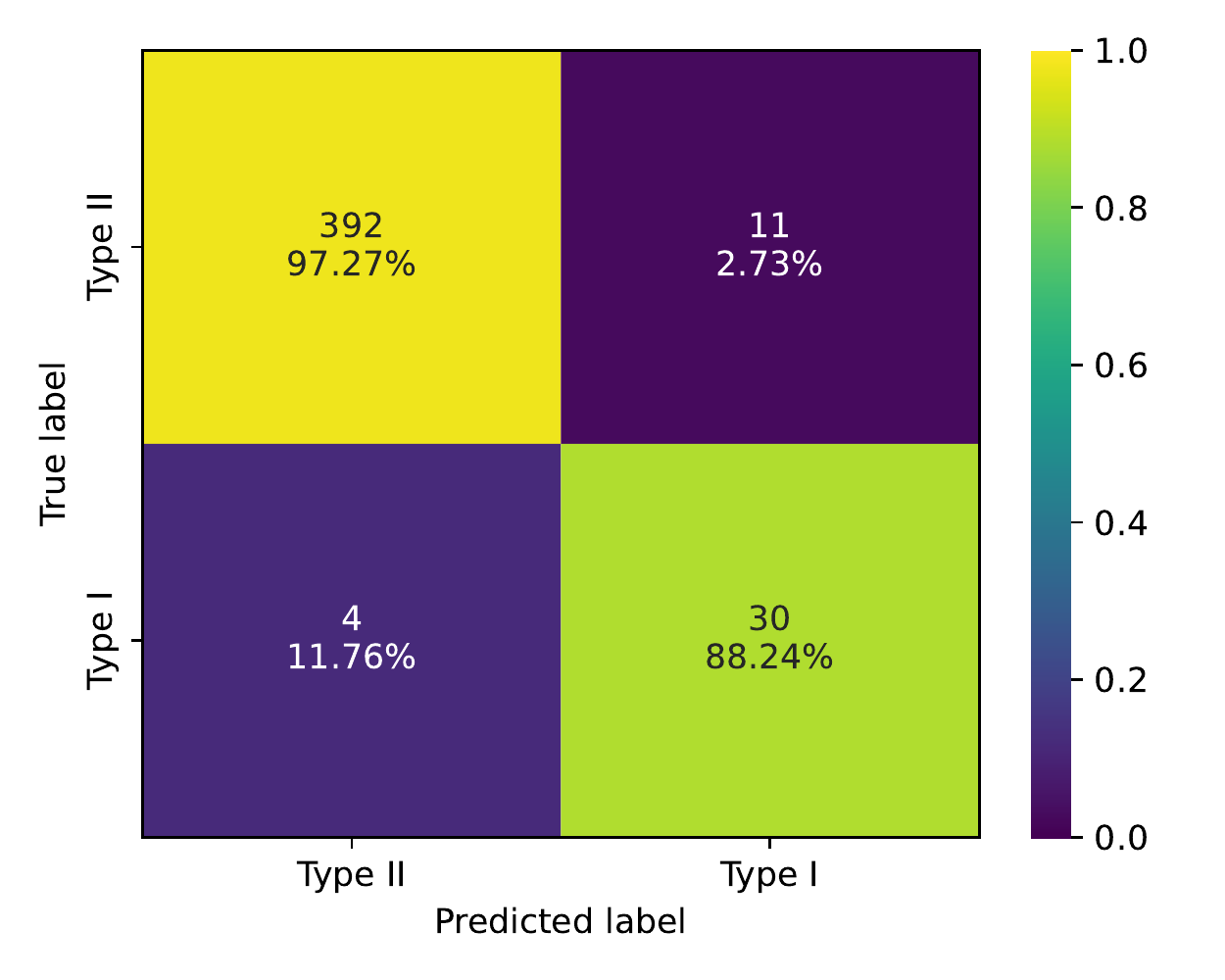}
    \label{fig:all_cm}}
\subfloat[Confusion matrix on intermingled GRBs in the test set]{
    \includegraphics[width=0.48\textwidth]{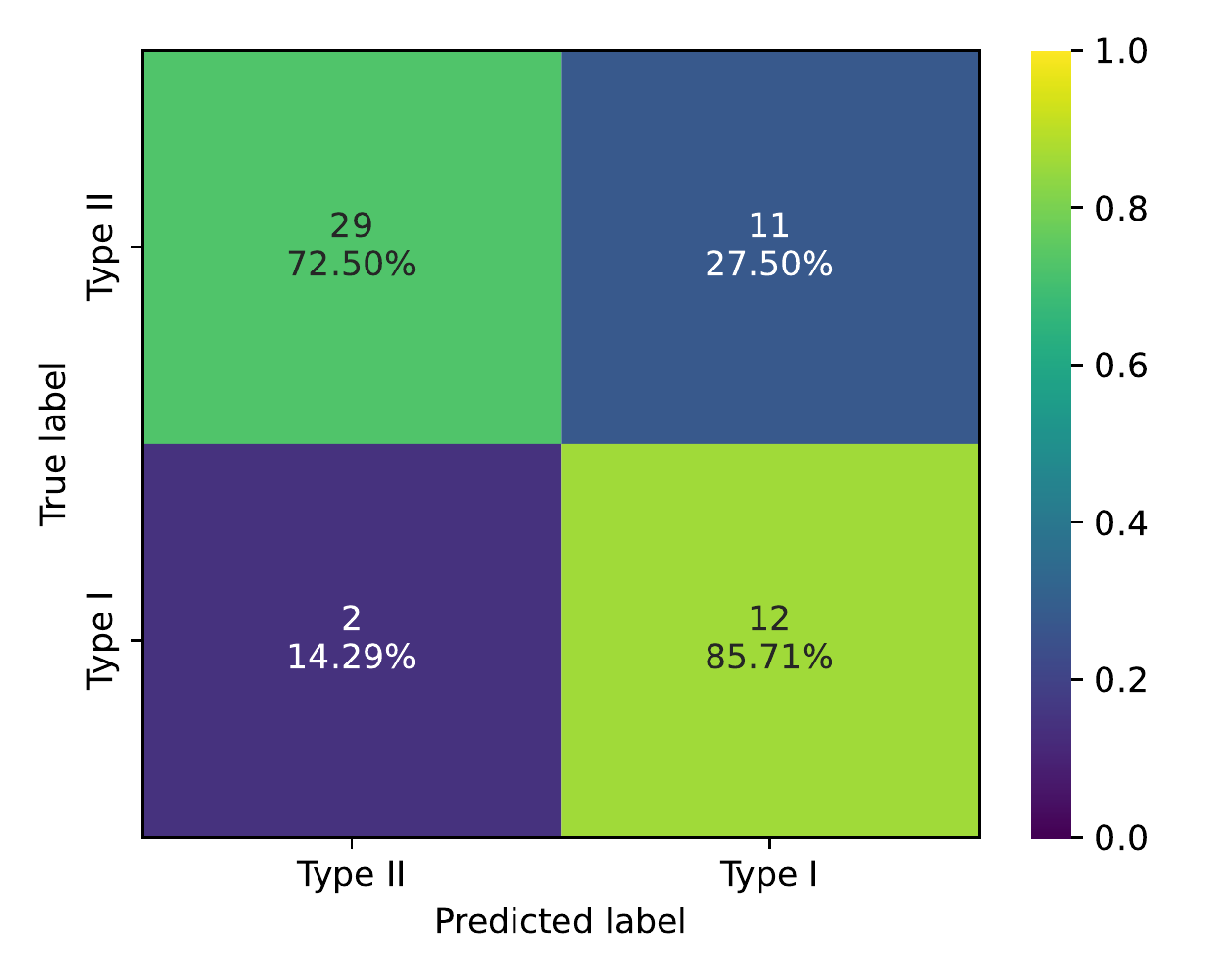}
    \label{fig:all_inter_cm}}\\
\subfloat[Confusion matrix on intermediate GRBs in the test set]{
    \includegraphics[width=0.48\textwidth]{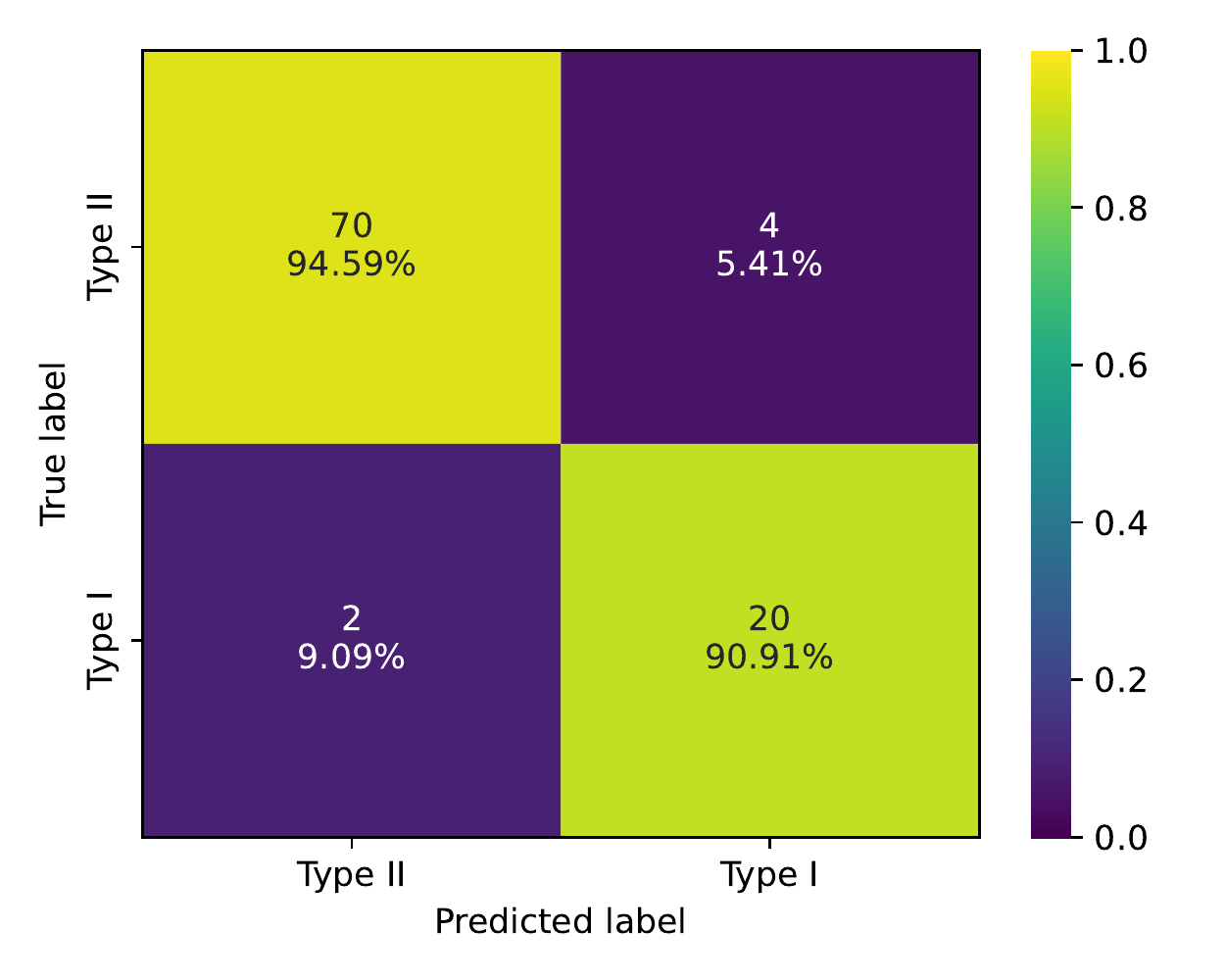}
    \label{fig:all_mediate_cm}}
\subfloat[Average SHAP values of each feature on the training set]{
    \includegraphics[width=0.48\textwidth]{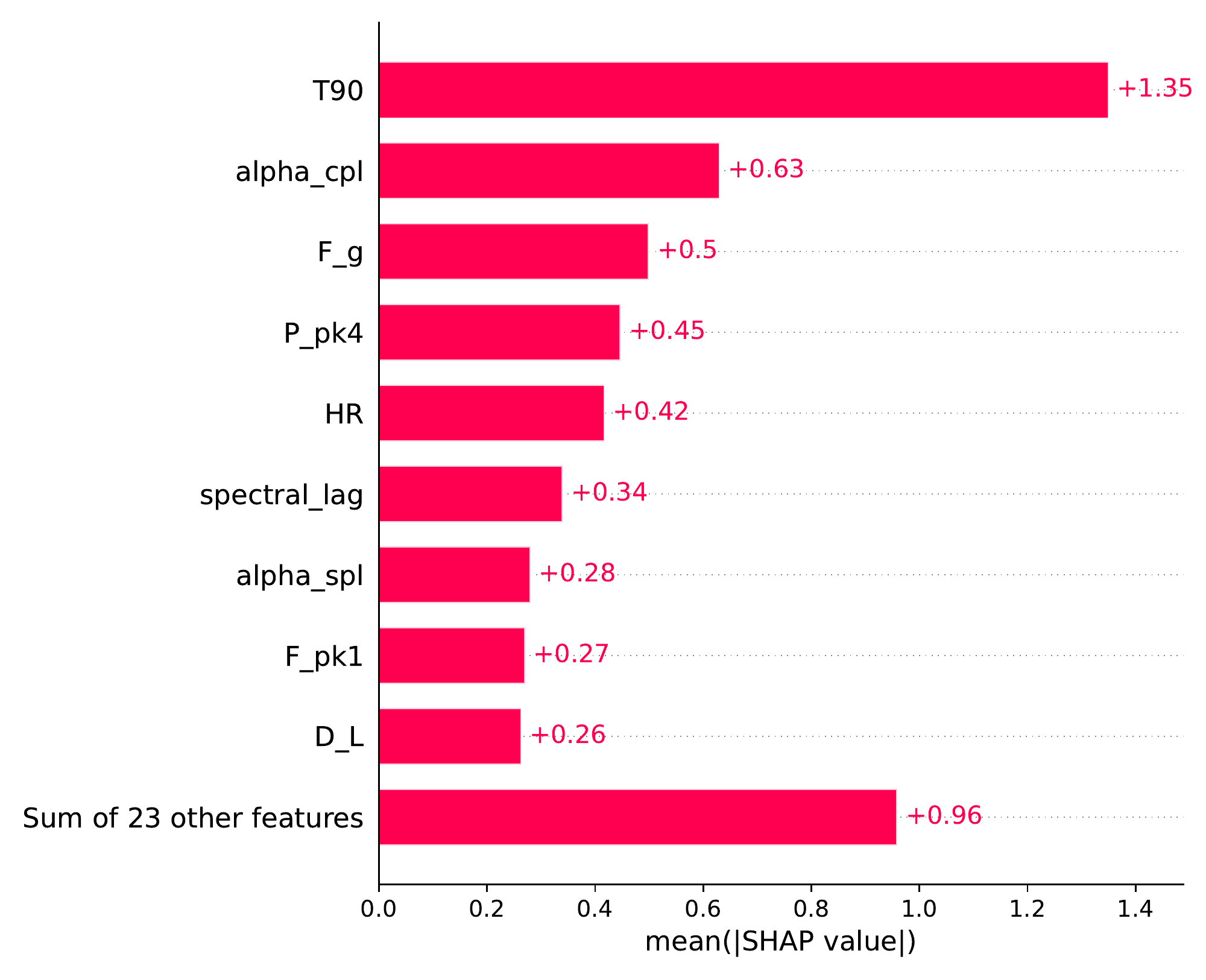}
    \label{fig:all_fi}}\\
\subfloat[SHAP value beeswarm plot on the training set]{
    \includegraphics[width=0.48\textwidth]{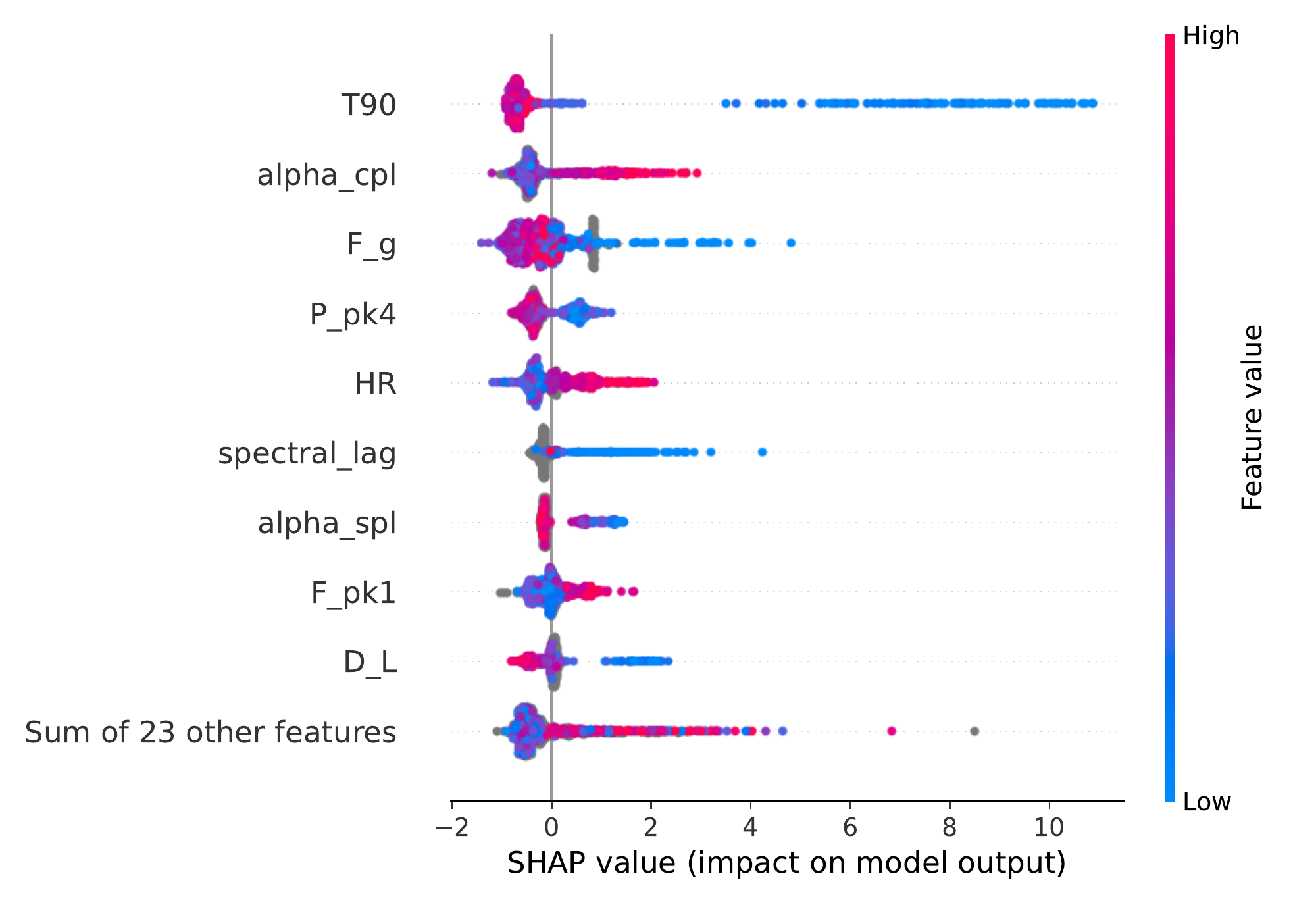}
    \label{fig:all_fi2}}
\caption{Examples of confusion matrices and SHAP feature importance values of the all features subgroup.}
\label{fig:all}
\end{figure*}

The corresponding confusion matrices and feature importance are shown in Figure \ref{fig:all}. The most important features all come from the prompt emission subgroup, which shows that prompt emission data is most important in GRB classification.

\subsection{Comparing the feature subgroups}
\label{subsec:compare_features}

Because the training and test set splitting process introduces randomness to the results, $F_1$ scores from a single trial may not be able to fully reflect the abilities in distinguishing Type I and II GRBs for different feature subgroups. Therefore, we repeat the random splitting and training process 1000 times, and record the $F_1$ scores of each feature subgroup on the entire test set and intermingled GRBs.

We report the average $F_1$ scores, along with standard deviations based on the 1000 trials for each feature subgroup on the entire test set and intermingled GRBs in Table \ref{table:f1_list}. We found that the prompt emission subgroup performs the best in predicting Type I and II GRBs, while the average $F_1$ score of the afterglow subgroup is significantly lower. Using all features only marginally improve the performance of the model. Host galaxy comes in between the two subgroups. However, prompt emission including $T_{90}$ performs the worst on the intermingled GRBs. Also, all the feature subgroups performs reasonably well on the intermediate GRBs, which indirectly rejects the existence of a third intermediate GRB type.

Note that the intermingled and intermediate GRBs form a smaller sample that usually get more attention from the scientific community compared with all the GRBs. Among the 32 features we use in this study, requiring the GRBs to have $T_{90}$ and one other feature to be known, the general GRB sample on average have 2.5 known features, while the intermingled and intermediate samples on average have 9.3 and 9.4 features to be known, respectively.

We also the compare the performance of our model with the traditional way of classifying GRBs on the T90-HR plane by building a decision tree \citep[e.g.][]{breiman1984ClassificationRegressionTrees,timofeev2004ClassificationRegressionTrees,loh2011ClassificationRegressionTrees,loh2014FiftyYearsClassification} with $T_{90}$ and hardness ratio as input. We use the implementation in \textsc{scikit-learn} and set the maximum depth of the decision tree to be 3. We use this model instead of the more sophisticated \textsc{XGBoost} model we use in the other parts of this study because we think the decision tree model better reflects the classification ability of a human scientist.

\begin{figure*}
\centering
\subfloat[Confusion matrix on all GRBs in the test set]{
    \includegraphics[width=0.49\textwidth]{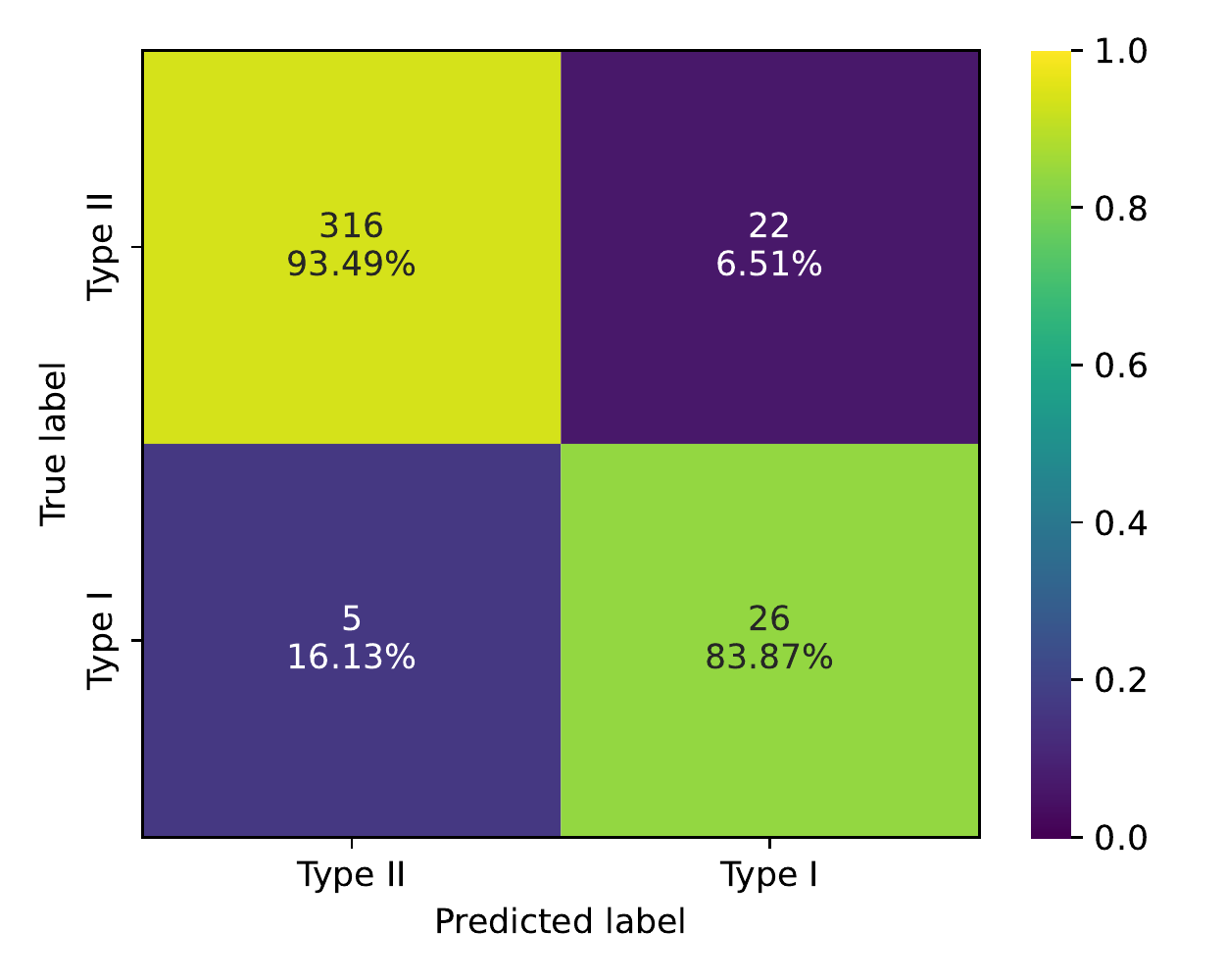}
    \label{fig:tree_cm}}
\subfloat[Confusion matrix on intermingled GRBs in the test set]{
    \includegraphics[width=0.49\textwidth]{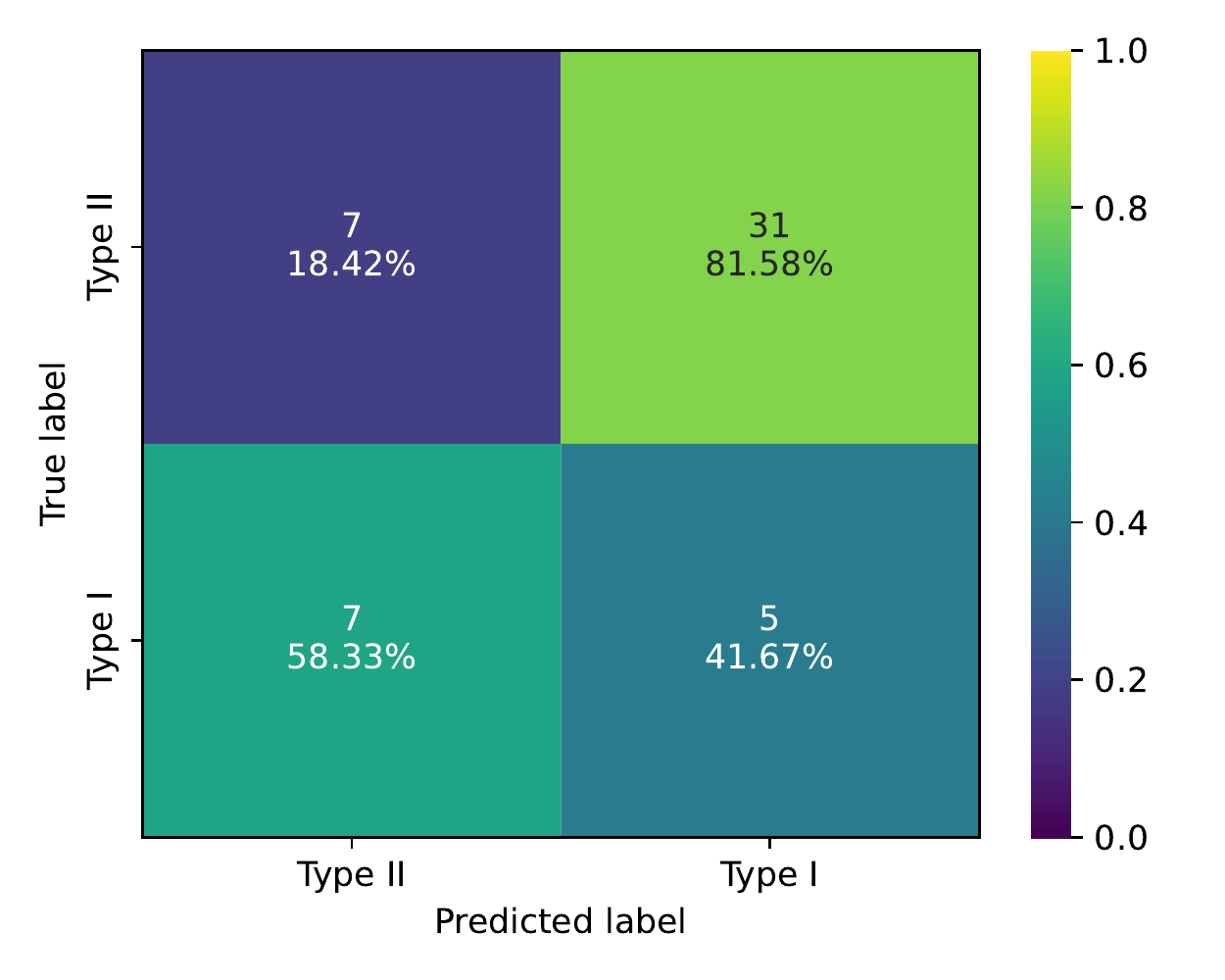}
    \label{fig:tree_inter_cm}}\\
\subfloat[Confusion matrix on intermediate GRBs in the test set]{
    \includegraphics[width=0.49\textwidth]{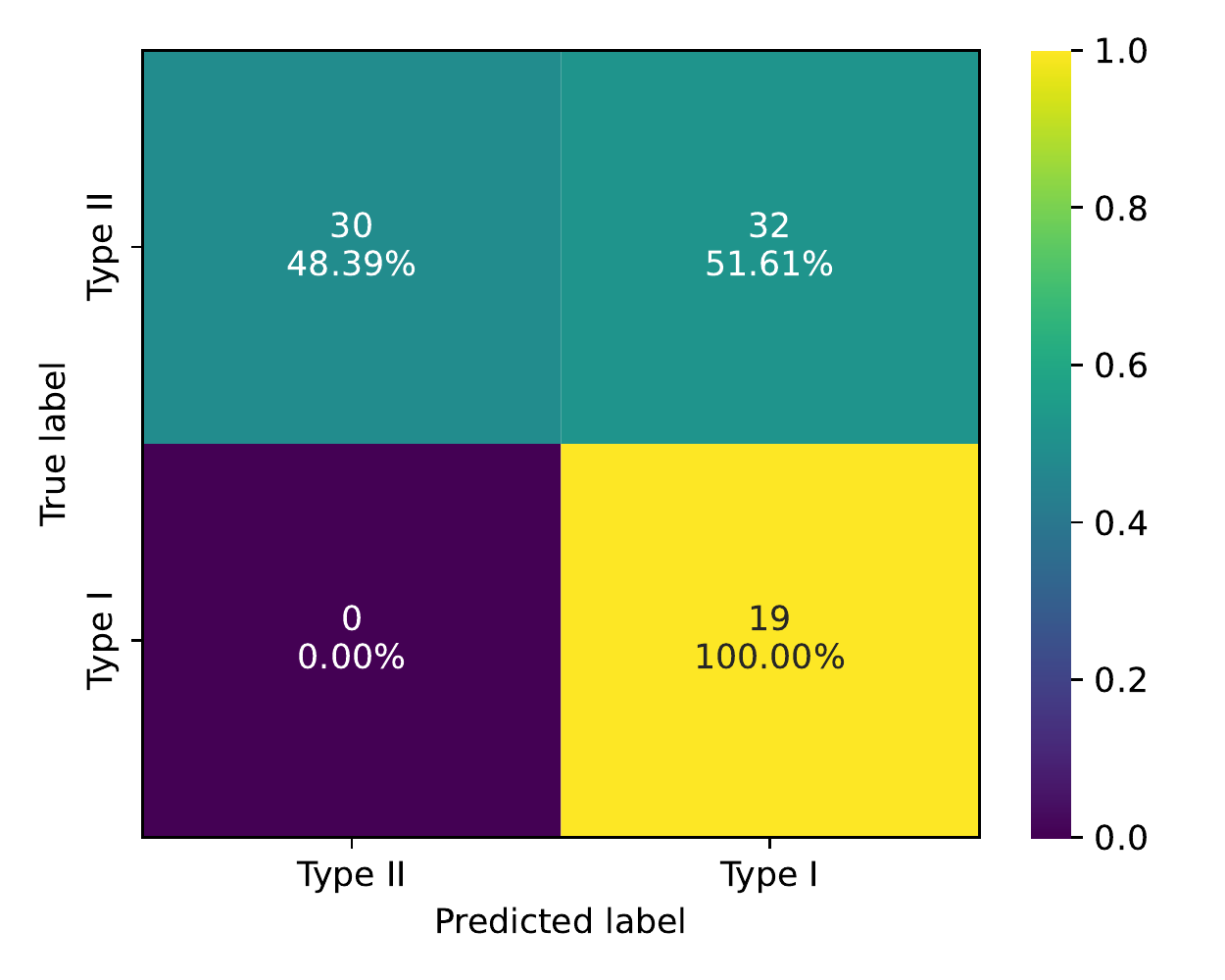}
    \label{fig:tree_mediate_cm}}
\caption{Examples of confusion matrices of the $T_{90}$--HR decision tree.}
\label{fig:tree}
\end{figure*}

With this decision tree model, we are able to achieve $F_1$ score of 0.761 on all GRBs, 0.125 on intermingled GRBs and 0.667 on intermediate GRBs. The examples of confusion matrices are shown in Figure \ref{fig:tree}. When comparing the average $F_1$ scores from multiple trials listed in Table \ref{table:f1_list}, we find that while the performance of our \textsc{XGBoost} multi-parameter model and the simple decision tree model are comparable on all the GRBs, the multi-parameter model performs significantly better on the intermingled and intermediate GRBs. This shows that our new classification method is an improvement over the traditional one, especially on the intermingled and intermediate GRBs.

\begin{table*}
    \begin{tabular}{ccccc}
    \hline
    Feature subgroup & All GRBs $F_1$ & Intermingled $F_1$ & Intermediate $F_1$\\
	\hline
	Prompt emission & $0.758\substack{+0.042 \\ -0.046}$ & $0.633\substack{+0.089 \\ -0.087}$ & $0.848\substack{+0.056 \\ -0.057}$\\
    Prompt emission, without \texttt{T90} & $0.665\substack{+0.057 \\ -0.059}$ & $0.755\substack{+0.072 \\ -0.070}$ & $0.869\substack{+0.040 \\ -0.048}$\\
    Prompt emission, without \texttt{T90}, \texttt{HR}, \texttt{F\_g} & $0.497\substack{+0.062 \\ -0.063}$ & $0.773\substack{+0.084 \\ -0.080}$ & $0.855\substack{+0.049 \\ -0.055}$\\
    Afterglow & $0.425\substack{+0.108 \\ -0.103}$ & $0.798\substack{+0.135 \\ -0.131}$ & $0.831\substack{+0.110 \\ -0.081}$\\
    Host galaxy & $0.641\substack{+0.079 \\ -0.081}$ & $0.909\substack{+0.091 \\ -0.076}$ & $0.941\substack{+0.059 \\ -0.066}$\\
    Host galaxy, without \texttt{offset} & $0.490\substack{+0.119 \\ -0.115}$ & $0.803\substack{+0.106 \\ -0.136}$ & $0.878\substack{+0.122 \\ -0.109}$\\
    All & $0.767\substack{+0.045 \\ -0.043}$ & $0.638\substack{+0.090 \\ -0.086}$ & $0.859\substack{+0.046 \\ -0.048}$\\
    $T_{90}$, HR decision tree & $0.757\substack{+0.049 \\ -0.044}$ & $0.144\substack{+0.052 \\ -0.024}$ & $0.636\substack{+0.044 \\ -0.068}$\\
	\hline
	\end{tabular}
	\caption{List of average $F_1$ scores and 16th/84th percentile percentile values obtained with different feature subgroups and GRB samples.}
	\label{table:f1_list}
\end{table*}

\section{Predicting unclassified GRBs}
\label{sec:prediction}

After building the models, we then move on to predict the classes of the unclassified GRBs in the Big Table. Since the all-feature subgroup achieved the best performance, we use all the features to train our model and predict the classes.

We train the model using the same method described in Section \ref{sec:data} with all the classified GRBs with at least one feature we intend to use and $T_{90}$ known, and use the trained model to predict the probabilities of the unclassified GRBs being either class. We also require the unclassified GRBs to have at least one feature and $T_{90}$ known. 1455 GRBs are used for training, and the class probabilities of 2809 unclassified GRBs are predicted. For each unclassified GRB, the class in which they are predicted with the highest probability is assigned as their class. 2181 GRBs are predicted as Type II, while 628 GRBs are predicted as Type I. The prediction results are listed in Table \ref{table:predictions}. We graph the probability distribution of the unclassified GRBs being Type II in Figure \ref{fig:predict_proba}. To compare our results with the traditional method of classifying GRBs on the $T_{90}$--Hardness ratio plane, we also plot our prediction results on the $T_{90}$--Hardness ratio plane in Figure \ref{fig:predict_t90_hr}.

\begin{figure}
\centering
	\includegraphics[width=0.49\textwidth]{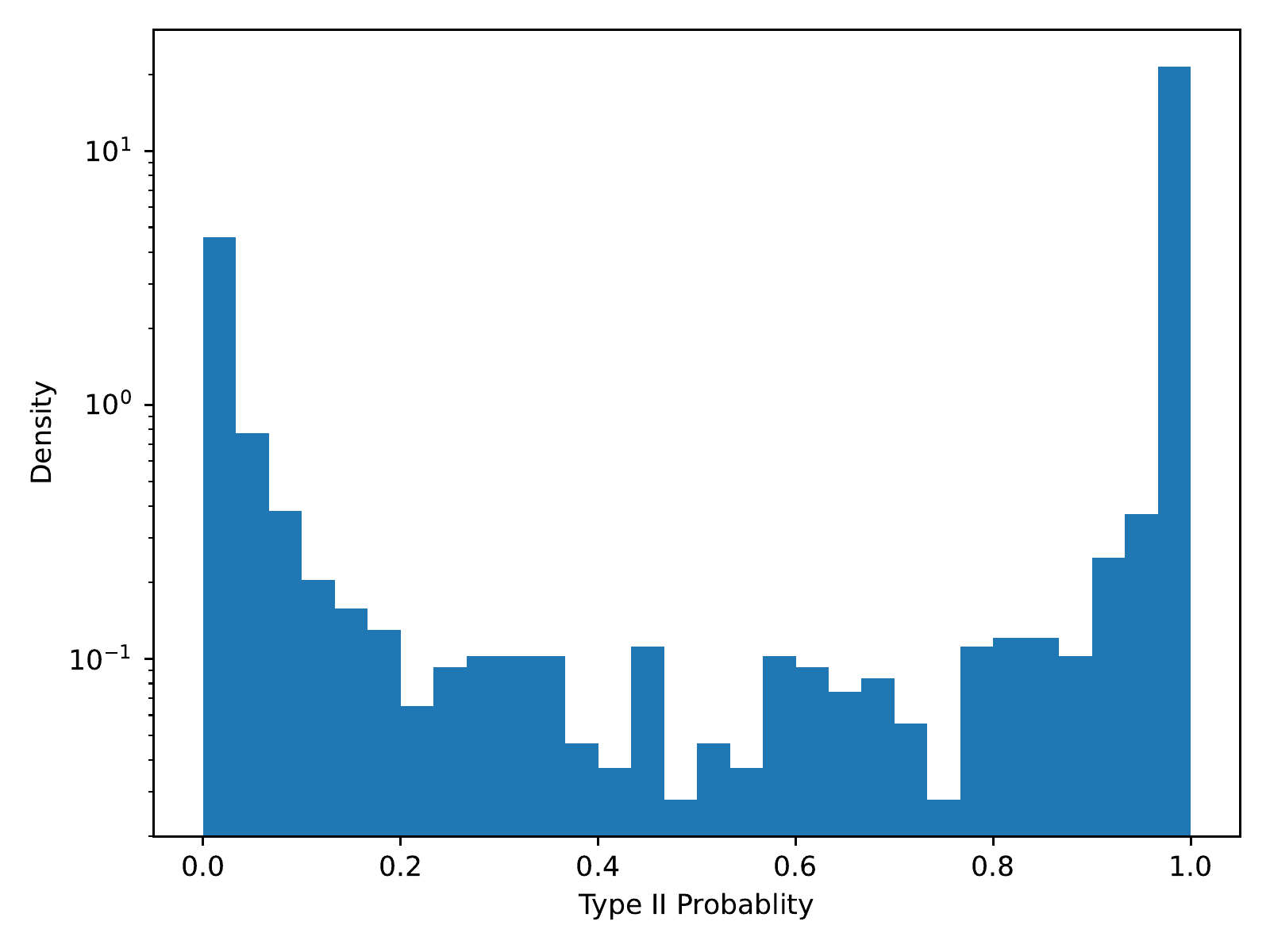}
	\caption{Probability distribution of the unclassified GRBs being Type II. The probability for Type I is $1-\mathrm{the}$ shown value.}
	\label{fig:predict_proba}
\end{figure}

\begin{figure}
\centering
	\includegraphics[width=0.49\textwidth]{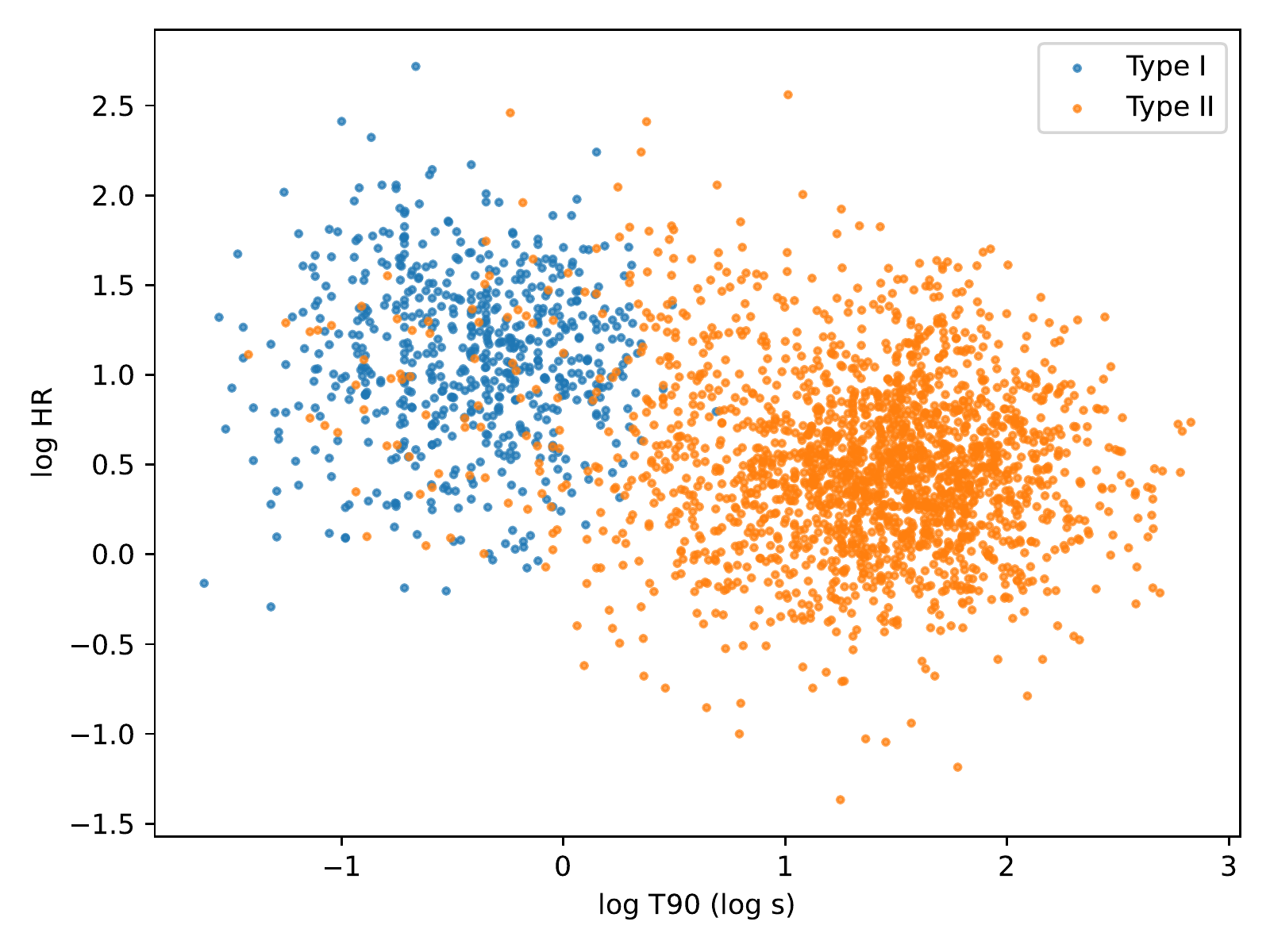}
	\caption{Distributions of \texttt{T90} and \texttt{HR} of the unclassified GRBs. The points are marked with their predicted classes.}
	\label{fig:predict_t90_hr}
\end{figure}

\begin{table}
    \centering
    \begin{tabular}{llll}
    \hline
    \multicolumn{1}{c}{GRB}&$p_{\mathrm{I}}$&$p_{\mathrm{II}}$&Type\\
	\hline
    170325B & 0.0 & 1.0 & II \\
    170315A & 0.0 & 1.0 & II \\
    170309A & 0.12 & 0.88 & II \\
    170228B & 0.0 & 1.0 & II \\
    170208C & 0.0 & 1.0 & II \\
    170207A & 0.0 & 1.0 & II \\
    170206C & 0.003 & 0.997 & II \\
    170130A & 0.0 & 1.0 & II \\
    170121A & 0.381 & 0.619 & II \\
    170120A & 0.0 & 1.0 & II \\
	\hline
	\end{tabular}
	\caption{Prediction results of the unclassified GRBs. The probability of them being Type I or II are shown as $p_{\mathrm{I}}$ and $p_{\mathrm{II}}$ respectively. This is an example of the first ten rows of the table. The full version is published in its entirety in the machine-readable format.}
	\label{table:predictions}
\end{table}

\section{Conclusions and Discussions}
\label{sec:conclusions}
In this paper, we applied supervised machine methods, mainly \textsc{XGBoost}, to the classification of Type I and II GRBs. We come up with the following conclusions:

\begin{itemize}
  \item Classifying GRBs solely based on $T_{90}$ can yield unsatisfactory results, especially on intermingled GRBs. Criteria based on multiple observational parameters are needed.
  \item Compared with traditional GRB classification methods, the machine learning method can effectively classify GRBs, especially intermingled and intermediate ones.
  \item The fact that supervised machine learning model with two classes of GRBs can effectively classify intermediate GRBs with $T_{90}$ between \SIrange{1}{4}{\s} indirectly rejects the existence of a third intermediate GRB class proposed based on duration distribution.
  \item We found that the best feature group in predicting Type I or II GRB is prompt emission. Among features on prompt emission, we found that $T_{90}$ still separates Type I and II GRBs the best. Besides $T_{90}$, fluence and hardness ratio are also important features. Since fluence is correlated with $T_{90}$, this is consistent with the traditional way of classifying GRBs on the $T_{90}$ -- HR plane.
  \item We predict the class of some of the GRBs not present in Greiner's catalog. Their predicted class and their probabilities of being either are shown in Table \ref{table:predictions}.
  \item The methods employed in this study can be applied to future newly discovered GRBs to identify potentially peculiar GRBs in their early stages and help allocate resources for follow-up observations. The analysis code used in this study is available at \url{https://github.com/Rigel7/grb-ml}.
\end{itemize}

Recently, three possible intermingled GRBs have gained a lot of attention from the scientific community. GRB 200826A is thought to be an intermingled Type II GRB \citep{zhang2021PeculiarlyShortdurationGammaray,ahumada2021DiscoveryConfirmationShortest,rossi2022PeculiarShortdurationGRB}, while GRB 211221A and GRB 230307A are thought to be intermingled Type I GRBs \citep{yang2022LongdurationGammarayBurst,rastinejad2022KilonovaFollowingLongduration,sun2023MagnetarEmergencePeculiar}. We apply our trained model to these three GRBs to examine their observational proprieties. The features we gathered and corresponding references for the three GRBs are listed in Table \ref{table:GRB200826A}, \ref{table:GRB211211A} and \ref{table:GRB230307A}.

\begin{figure*}
\centering
\includegraphics[width=\textwidth]{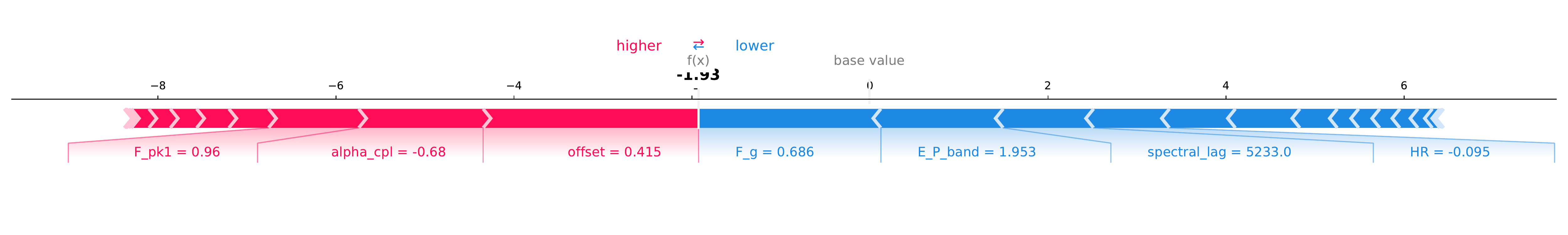}
\includegraphics[width=\textwidth]{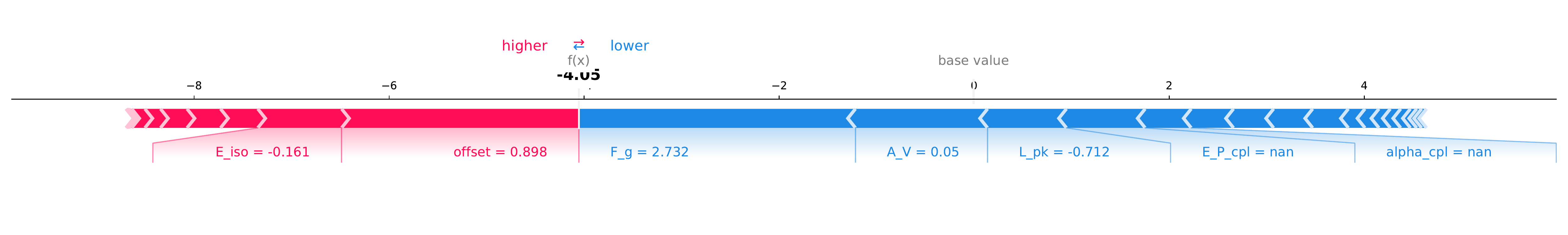}
\includegraphics[width=\textwidth]{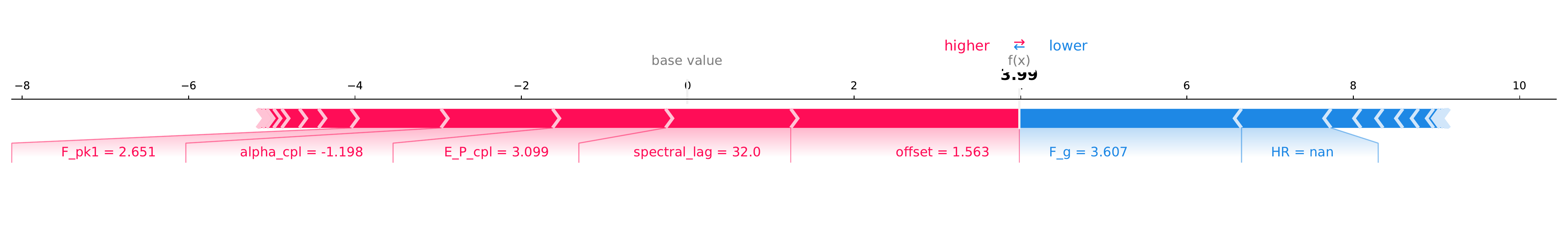}
\caption{SHAP values of individual feature values of GRB 200826A, GRB 211211A and GRB 230307A. Features marked with red color push the prediction results toward Type I, while features marked with blue color push the prediction results toward Type II. Upper: GRB 200826A. Middle:GRB 211211A. Lower: GRB 230307A.}
\label{fig:new_grb}
\end{figure*}

Since $T_{90}$ can be misleading for the classification of intermingled GRBs, we use all the features except $T_{90}$ to classify these three bursts. For GRB 200826A, the model predicts it to have 13\% probability of being Type I and 87\% probability of being Type II. For GRB 211211A, the model predicts it to have 2\% probability of being Type I and 98\% probability of being Type II. For GRB 230307A, the model predicts it to have 98\% probability of being Type I and 2\% probability of being Type II. The SHAP values explaining how the individual features affect the predictions results are shown in Figure \ref{fig:new_grb}.

The prediction results of GRB 200826A and GRB 230307A from our model match the proposed classification of the two GRBs by other literature, while the prediction result of GRB 211211A matches the classification done solely based on its duration. This shows that these GRBs, particularly GRB 211211A are truly peculiar and more study should be done on their observational properties and physical origins. Indeed, \citet{yang2022LongdurationGammarayBurst} proposes that GRB 211211A could be originated from a white dwarf -- neutron star merger, while \citet{barnes2023CollapsarOriginGRB} argue that GRB 211211A could still be explained by the normal Type II collapsar GRB model.

With its ability to not only predict the physical types of GRBs, but also explain the importance of each parameter in individual classifications, our model can provide independent opinions on the classifications of possible peculiar GRBs and help guide future observations and studies.

\begin{table}
    \centering
    \addtolength{\tabcolsep}{-3pt}
    \begin{tabular}{llll}
    \hline
    Name & Value & Unit & Reference\\
	\hline
    \texttt{T90} & 1.14 & \si{\s} & GCN 28287\\
    \texttt{F\_g} & 4.85 & $10^{-6}\;\si{\erg\per\centi\m\squared}$ & \citet{zhang2021PeculiarlyShortdurationGammaray}\\
    \texttt{HR} & 0.803 &  --- & \citet{zhang2021PeculiarlyShortdurationGammaray}\\
    \texttt{F\_pk1} & 9.11 & $10^{-6}\;\si{\erg\per\centi\m\squared\per\s}$ & \citet{zhang2021PeculiarlyShortdurationGammaray}\\
    \texttt{P\_pk4} & 39.06 & \si{\photon\per\centi\m\squared\per\s} & GCN 28287\\
    \texttt{alpha\_band} & -0.41 & --- & GCN 28287\\
    \texttt{beta\_band} & -2.4 & --- & GCN 28287\\
    \texttt{E\_P\_band} & 89.8 & $\si{\kilo\eV}$ & GCN 28287\\
    \texttt{alpha\_cpl} & -0.68 & --- & \citet{zhang2021PeculiarlyShortdurationGammaray}\\
    \texttt{E\_P\_cpl} & 120.29 & $\si{\kilo\eV}$ & \citet{zhang2021PeculiarlyShortdurationGammaray}\\
    \texttt{spectral\_lag} & 5233 & \si{\milli\s\per\mega\eV} & \citet{zhang2021PeculiarlyShortdurationGammaray}\\
    \hline
    \texttt{z} & 0.714 & --- & GCN 28301\\
    \texttt{D\_L} & 1.21 & $10^{28}\;\mathrm{cm}$ & GCN 28301\\
    \texttt{E\_iso} & 0.709 & $10^{52}\;\si{\erg}$ & \citet{zhang2021PeculiarlyShortdurationGammaray}\\
    \texttt{L\_pk} & 1.41 & $10^{52}\;\si{\erg\per\s}$ & \citet{zhang2021PeculiarlyShortdurationGammaray}\\
	\hline
    \texttt{offset} & 2.6 & \si{\kilo\parsec} & \citet{zhang2021PeculiarlyShortdurationGammaray}\\
    \texttt{metallicity} & -0.37 & --- & \citet{gupta2022PhotometricStudiesHost}\\
    \texttt{A\_V} & 0.19 & --- & \citet{gupta2022PhotometricStudiesHost}\\
    \texttt{SFR} & 3.49 & $\si{\Msun\per\year}$& \citet{gupta2022PhotometricStudiesHost}\\
    \texttt{Age} & 4.74 & \si{\mega\year} & \citet{gupta2022PhotometricStudiesHost}\\
    \texttt{Mass} & $8.32\times10^9$ & $\si{\Msun}$ & \citet{gupta2022PhotometricStudiesHost}\\
	\hline
	\end{tabular}
	\addtolength{\tabcolsep}{3pt}
	\caption{Observational properties of GRB 200826A.}
	\label{table:GRB200826A}
\end{table}

\begin{table}
    \centering
    \addtolength{\tabcolsep}{-3pt}
    \begin{tabular}{llll}
    \hline
    Name & Value & Unit & Reference\\
	\hline
    \texttt{T90} & 34.3 & \si{\s} & GCN 31210\\
    \texttt{F\_g} & 540 & $10^{-6}\;\si{\erg\per\centi\m\squared}$ & GCN 31223\\
    \texttt{HR} & 3.6 &  --- & GCN 31209, GCN 31223\\
    \texttt{P\_pk4} & 324.9 & \si{\photon\per\centi\m\squared\per\s} & GCN 31210\\
    \texttt{alpha\_band} & 1.3 & --- & GCN 31210\\
    \texttt{beta\_band} & 2.4 & --- & GCN 31210\\
    \texttt{E\_P\_band} & 646.8 & $\si{\kilo\eV}$ & GCN 31210\\
    \texttt{alpha\_spl} & 1.56 & --- & GCN 31209\\
    \texttt{spectral\_lag} & 107 & \si{\milli\s\per\mega\eV} & \citet{rastinejad2022KilonovaFollowingLongduration}\\
    \hline
    \texttt{z} & 0.0763 & --- & GCN 31221\\
    \texttt{D\_L} & 0.105 & $10^{28}\;\mathrm{cm}$ & GCN 31221\\
    \texttt{E\_iso} & 0.69 & $10^{52}\;\si{\erg}$ & GCN 31223\\
    \texttt{L\_pk} & 0.194 & $10^{52}\;\si{\erg\per\s}$ & \citet{yang2022LongdurationGammarayBurst}\\
	\hline
    \texttt{offset} & 7.91 & \si{\kilo\parsec} & \citet{rastinejad2022KilonovaFollowingLongduration}\\
    \texttt{metallicity} & -0.69 & --- & \citet{rastinejad2022KilonovaFollowingLongduration}\\
    \texttt{A\_V} & 0.05 & --- & \citet{rastinejad2022KilonovaFollowingLongduration}\\
    \texttt{Age} & 4.00 & \si{\mega\year} & \citet{rastinejad2022KilonovaFollowingLongduration}\\
    \texttt{Mass} & $6.92\times10^8$ & $\si{\Msun}$ & \citet{rastinejad2022KilonovaFollowingLongduration}\\
	\hline
	\end{tabular}
	\addtolength{\tabcolsep}{3pt}
	\caption{Observational properties of GRB 211211A.}
	\label{table:GRB211211A}
\end{table}

\begin{table}
    \centering
    \addtolength{\tabcolsep}{-3pt}
    \begin{tabular}{llll}
    \hline
    Name & Value & Unit & Reference\\
	\hline
    \texttt{T90} & 41.52 & \si{\s} & \citet{sun2023MagnetarEmergencePeculiar}\\
    \texttt{F\_g} & 4050 & $10^{-6}\;\si{\erg\per\centi\m\squared}$ & GCN 33579\\
    \texttt{F\_pk1} & 448 & $10^{-6}\;\si{\erg\per\centi\m\squared\per\s}$ & \citet{sun2023MagnetarEmergencePeculiar}\\
    \texttt{P\_pk4} & 791 & \si{\photon\per\centi\m\squared\per\s} & GCN 33411\\
    \texttt{alpha\_band} & -0.43 & --- & GCN 33444\\
    \texttt{beta\_band} &-5.1 & --- & GCN 33444\\
    \texttt{E\_P\_band} & 1260 & $\si{\kilo\eV}$ & GCN 33444\\
    \texttt{alpha\_cpl} & -1.198 & --- & \citet{sun2023MagnetarEmergencePeculiar}\\
    \texttt{E\_P\_cpl} & 1254.68 & $\si{\kilo\eV}$ & \citet{sun2023MagnetarEmergencePeculiar}\\
    \texttt{spectral\_lag} & 32 & \si{\milli\s\per\mega\eV} & \citet{sun2023MagnetarEmergencePeculiar}\\
    \hline
    \texttt{z} & 0.065 & --- & \citet{sun2023MagnetarEmergencePeculiar}\\
    \texttt{D\_L} & 0.0907 & $10^{28}\;\mathrm{cm}$ & \citet{sun2023MagnetarEmergencePeculiar}\\
    \texttt{E\_iso} & 3.08 & $10^{52}\;\si{\erg}$ & \citet{sun2023MagnetarEmergencePeculiar}\\
    \texttt{L\_pk} & 48.9 & $10^{52}\;\si{\erg\per\s}$ & \citet{sun2023MagnetarEmergencePeculiar}\\
	\hline
    \texttt{offset} & 36.6 & \si{\kilo\parsec} & \citet{sun2023MagnetarEmergencePeculiar}\\
    \texttt{metallicity} & 8.2 & --- & \citet{levan2023JWSTDetectionHeavy}\\
    \texttt{SFR} & 0.547 & $\si{\Msun\per\year}$& \citet{levan2023JWSTDetectionHeavy}\\
    \texttt{Age} & 1130 & \si{\mega\year} & \citet{levan2023JWSTDetectionHeavy}\\
    \texttt{Mass} & $1.65\times10^9$ & $\si{\Msun}$ & \citet{levan2023JWSTDetectionHeavy}\\
	\hline
	\end{tabular}
	\addtolength{\tabcolsep}{3pt}
	\caption{Observational properties of GRB 230307A.}
	\label{table:GRB230307A}
\end{table}

\begin{acknowledgments}
This work is partially supported by the Top Tier Doctoral Graduate Research Assistantship (TTDGRA) and Nevada Center for Astrophysics at the University of Nevada, Las Vegas.
\end{acknowledgments}

\bibliographystyle{aasjournal}
\bibliography{grb-ml}
\end{document}